\documentclass[]{aa} 

\usepackage[colorlinks=true,linkcolor=blue, urlcolor=blue, citecolor=blue]{hyperref}

\usepackage{natbib}
\bibpunct{(}{)}{;}{a}{}{,} 

\usepackage{graphicx}

\usepackage[toc, page]{appendix}

\usepackage{multirow}

\usepackage{txfonts}

\begin{document}

   \title{Robust Binding Energy Distribution Sampling on Amorphous Solid Water Models}

   \subtitle{Method testing and validation with NH$_3$, CO and CH$_4$}

   \author{Maria Groyne\inst{1,2} \and Benoît Champagne\inst{2} \and Cedric Baijot\inst{1} \and Michaël De Becker\inst{1}}

   \institute{Space Sciences, Technologies and Astrophysics Research (STAR) Institute, University of Liège, Quartier Agora, 19c, Allée du 6 Aôut, B5c, B-4000 Sart Tilman, Belgium\\            
   \and Theoretical Chemistry Lab, Unit of Theoretical and Structural Physical Chemistry, Namur Institute of Structured Matter, University of Namur, Rue de Bruxelles, 61, B-5000 Namur, Belgium\\}

   \date{Received 9 April 2025 / Accepted 25 April 2025}
 
  \abstract 
   {The astrochemically efficient icy mantles surrounding dust grains in molecular clouds have been shown to be of a water-rich amorphous nature. This therefore implies a distribution of binding energies (BE) per species instead of a single value. Methods proposed so far for inferring BE and their distributions on amorphous ices rely on different approaches and approximations, leading to disparate results or BE dispersions with partially overlapping ranges.}
   {This work aims to develop a method based on a structurally reliable ice model and a statistically and physico-chemically robust approach for BE distribution inference, with the aim to be applicable to various relevant interstellar species.}
   {A multiscale computational approach is presented, with a Molecular Dynamics (MD) Heat \& Quench protocol for the amorphous water ice model, and an ONIOM(B3LYP-D3(BJ)/6-311+G**:GFN2-xtb) scheme for the BE inference, with a prime emphasis onto the BE/real system size convergence. The sampling of the binding configurations is twofold, exploring both regularly spaced binding sites, as well as various adsorbate-to-substrate orientations on each locally distinct site. This second source of BE diversity accounts for the local roughness of the potential energy landscape of the substrate. Three different adsorbate test cases are considered, i.e. NH$_3$, CO and CH$_4$, owing to their significance in dust icy mantles, and their distinct binding behavior with water ices. }
   {The BE distributions for NH$_3$, CO and CH$_4$ have been inferred, with converged statistics. The distribution for NH$_3$ is better represented by a double Gaussian component profile. Three starting adsorbate orientations per site are required to reach convergence for both Gaussian components of NH$_3$, while 2 orientations are sufficient for CO, and one unique for CH$_4$ (symmetric). Further geometrical and molecular surrounding insights have been provided. These results encompass previously reported results.  }
   {}

   \keywords{Astrochemistry, ISM: clouds, ISM:molecules, Molecular data, Methods: numerical, Methods:statistical}

   \maketitle

\section{Introduction}
\label{sec:Intro}
   
   Dense molecular clouds and young protostellar cores are known to be efficient astrochemical factories. This however cannot be solely attributed to the gas phase chemistry, with densities typically ranging in the [10$^3$-10$^6$] particles/cm$^3$ range, and cryogenic T conditions (10-20 K). The gas phase chemistry therefore presents slow kinetics and is dominated by exothermic (or weakly endothermic) processes, generally leading to fragmented products. The gas-phase astrochemistry taking place thereby faces difficulties in climbing the molecular complexity scale. 

   On the other hand, interstellar dust grains constitute a minor, i.e. representing only $\sim$ 1\% in mass with respect to the gas phase, but very important ingredient in such clouds. In fact, thanks to their surface ability to act either as a passive third body taking away the excess of reaction energy, an active chemical catalyst, or a hub keeping in close proximity reaction partners for eventual subsequent reactions, dust grains facilitate the ascent of the molecular complexity scale. Among others, water molecules are formed in situ, through exothermic quantum-tunneling-mediated surface reactions \citep[]{Tielens&Hagen,Miyauchi,Dulieu,Oba,Molpeceres}. In that context, the conditions prevailing in the deep interior of such clouds favor the growth of a thick water-dominated icy mantle around the refractory grain core \citep{Oberg,Boogert,McClure}, exhibiting analogous hetero-catalytic properties as grain surfaces. More specifically, it is well recognized that molecular cloud and young protostellar core ices present a dominance of amorphous water-rich ices, generally referred as Amorphous Solid Water (ASW), as evidenced by near-infrared (NIR) spectral data \citep{Smith}. Furthermore, in parallel to H$_2$O formation, other species may adsorb on the surface, diffuse, possibly react, and eventually desorb. This solid phase chemistry is for instance needed to explain the abundance of some relevant interstellar species, starting with H$_2$ \cite[]{Gould&Salpeter,Hollenbach&Salpeter,Watson&Salpeter,Tielens&Hagen}, passing by CO$_2$, H$_2$CO, CH$_3$OH, ... \citep[e.g.][]{Fuchs,Ioppolo,Pirim&Krim,Minissale}, and is intrinsically linked to the binding interaction of the adsorbate to its substrate.

In that context, binding energies (BE) are key parameters for an accurate astrochemical modeling. Nevertheless, one of the main current issues in this field concerns the lack of accuracy in the knowledge of binding parameters (e.g. \cite{Minissale_review} and references therein), i.e. binding energy, diffusion energy and pre-exponential factor values. This is therefore a key limitation for the improvement of the accuracy of astronomical models. Indeed, considering that the desorption and diffusion processes are parameterized as Arrhenius-like functions, the exponential dependency on these energies and their associated uncertainty can lead to large discrepancies, as highlighted by the sensitivity analyses from \cite{Penteado} and \cite{Grassi}.
   
In terms of published works, several studies focused on BE inference on interstellar ice analogues. On the experimental aspect, BEs can be inferred from  Temperature Programmed Desorption (TPD) experiments \citep[e.g.][]{Noble}. However, this method is not suitable for volatile species due to sensitivity issues, nor for radicals, owing to their too short life-time. On the other hand, theoretical computational chemistry studies can in practice overcome such limitations. Nonetheless, focusing on ASW, no clear consensus on its structure has been approved, and completely different models are used. One may firstly cite the study from \cite{Wakelam}, focusing on the interactions of dimers formed by an adsorbate and a single water molecule, followed by a subsequent fit to empirical data. Others proposed the BE computation on small water clusters, such as in \cite{Das}, with very small H$_2$O assemblies, i.e. up to 6 molecules, or in \cite{Shimonishi} with clusters of 20 water molecules built from Molecular Dynamics (MD) simulations. Each of these three studies however suffers from two main issues, i.e. (i) the lack of consideration of proper long-range effects, including H-bond cooperativity, as well as undesired edge-effects from small clusters, and (ii) the inference of unique BE value per species. Nevertheless, ASW surfaces are expected to display a great diversity of structural arrangements due to the loss of long-range order, resulting in a complex substrate typology and a larger diversity of binding sites as compared to a crystalline phase. Hence, this implies a distribution of BEs. Some recent papers focus on the determination of such BE distributions on ASW surfaces, such as in e.g.\cite{Bovolenta} studying clusters of 22 water molecules. One may also cite \cite{Karssemeijer,Song_Kastner,Molpeceres_machine_learned_potential,Molpeceres_Kastner,Duflot,Ferrero} and \cite{Tinacci}, using larger clusters, studied through MD, Adaptive Kinetic Monte Carlo, periodic or multi-level/hybrid (Quantum Mechanics (QM)/Molecular Mechanics (MM), QM/QM' or ONIOM) computations. Most current astrochemical codes still rely on a single BE per species, but recent studies have proposed frameworks to include such distributions in kinetic models, such as in \cite{Grassi, Furuya}. This is very promising for the improvement of the solid-phase treatment in astrochemical codes and further supports the need for robust BE distributions to feed them.

In the scope of this paper, the main focus has consequently been directed toward the design of a BE distribution inference method with a robust ice model, both in terms of size and structure, while properly considering the statistical sampling of the binding sites. In that context, a model of ASW surface has been built through MD simulations, comprising 2000 water molecules. After verification of the reliability of the ice models through confrontation to empirical and other previously reported structural data, 100 regularly spaced sub-clusters are extracted for subsequent binding configuration sampling. On each sub-cluster, ONIOM-2 structure optimization and vibrational frequency computations on the ice/adsorbate adducts and their isolated ice counterpart are performed. This allows for the building of the BE distribution of relevant interstellar species. Note that three different adsorbate test cases will be studied throughout this study, namely NH$_3$, CO and CH$_4$. With the aim of building a method suitable to most relevant interstellar molecules, these have been chosen based on their distinct typical interactions they are forming with a water surface. 

In terms of contributions to the current state-of-the-art, the presented method provides insights into several issues that have not been (sufficiently) addressed in previous studies. Indeed: 

    \begin{itemize}
        \item[(i)] The structural reliability of the modeled ice has been deeply investigated through comparison to empirical structural data; 

        \item[(ii)] Concerning the design and optimization of the applied ONIOM scheme, the influence of system size on the computed BEs has been investigated through a benchmarking approach on the real system size;

        \item[(iii)] The BE distribution sampling procedure is twofold, including the exploration of the local roughness of the energy landscape of the substrate by probing different starting adsorbate-to-substrate orientations depending on the adsorbate symmetry itself. This introduces a second source of diversity in binding interactions, complementing the variety arising from the ice amorphicity and its complex surface typology;

        \item[(iv)] Considerable attention was devoted to ensure the convergence of the derived distributions with respect to their global statistics. Such convergence analyses were not implemented in the aforementioned papers, but should be considered to ensure the inference of statistically robust BE distributions. 
    \end{itemize}
   
The paper is divided as follows. Sect.\,\ref{sec:Methods} describes the methods for the ice model building and for the design of the BE computation scheme, as well as the foundations for the twofold binding configuration sampling. Sect.\,\ref{sec:Results} presents the results for the BE distributions built from the designed model, along with preliminary comparison to previously reported results. Sect.\,\ref{sec:Discussions} provides further insights onto the convergence of the mean and standard deviation of the inferred distributions, the binding behavior of each adsorbate and the astrochemical implication of this work. Finally, the main conclusions are summarized in Sect.\,\ref{sec:Conclusions}.

\section{modeling methods}
\label{sec:Methods}

\subsection{Ice model building}

The ice model has been prepared through MD simulations within a Heat \& Quench protocol. More specifically, an initial cubic box containing 2000 non-overlapping water molecules (TIP4P/2005 \citealt[]{TIP4P/2005}) has been constructed with the Packmol package \citep{PACKMOL}. The latter allows for the building of initial configurations for subsequent MD simulations by packing species in a defined space volume. The box size was chosen to match the density characteristics of ASW ice, reaching a density of $\sim$ 0.94 g/cm$^3$ for its most commonly discussed form, the i.e. Low Density Amorphous (LDA) phase. It resulted in a box of 40x40x40 \AA. The box is then submitted to MD simulation via the NAMD 2.14 software \citep{NAMD}\footnote{\url{http://www.ks.uiuc.edu/Research/namd/}}. The system is treated in the canonical NVT ensemble through the use of a Langevin thermostat. For the first part of the simulation, Periodic Boundary Conditions (PBC) are applied along the three Cartesian coordinates. Note also that the short-range non-bonded interactions are truncated at a cutoff distance of 12 \AA, with a switching distance and pair list distance of 8 and 14 $\AA$, respectively. The system is then equilibrated at 300 K for 10 ps. This equilibration at room temperature, reminiscent of liquid water, is aimed at ensuring disorder in the system for the subsequent quenching toward an amorphous phase. It is worth to mention that a ten times longer equilibration time has been tested, but has no influence on the temperature/energy variation amplitudes. The system is then annealed to 40 K, reminiscent of the LDA form of ASW. This annealing is achieved by sequential rapid temperature drops, rather than an instantaneous setting of the temperature to the targeted low value. This allows for a proper simulation of the phase transition. A Langevin dynamics is therefore applied at each 10 K temperature step for 1 ps (1000 timesteps of 1 fs), resulting in a temperature ramp of 10 K/ps. Note that different speeds of temperature ramps have been tested (e.g. 2K/ps), but it has no noticeable effects on the final structure based on Radial Distribution Functions (RDF) and cumulative running numbers, as well as H-bond analysis. At the end of this first quenching phase, the system is left for equilibration for another 10 ps. After this step, an intermediate check of the reliability of the structure of the modeled bulk ice has been performed, as commented hereafter. Actually, at this point of the ice model building procedure, the bulk structure of the ice has been modeled through the application of the PBC in the three space directions. In order to simulate a surface, the PBC in the z-direction has been artificially removed by adding a slab of void (100 \AA\hspace{0.05cm} thickness) above and below the box in the z direction. This way, PBC are conserved in the xy plane, while the molecules in the extreme z positions, i.e. top and bottom first layers of the simulated boxes, facing void, relax and reorient themselves from the typical bulk arrangement toward the ultimate definition of infinite surfaces. For this relaxation to take place, after another equilibration at the previously quenched temperature for 10 ps in the newly defined PBC box, the full system is re-heated to 100 K with a 2 K/ps temperature ramp, equilibrated for 10 ps at 100 K, re-quenched to 40 K with the same temperature ramp of 2 K/ps, and finally equilibrated at 40K for 10 additional ps.

\begin{figure}
    \centering
    \includegraphics[width=0.9\linewidth]{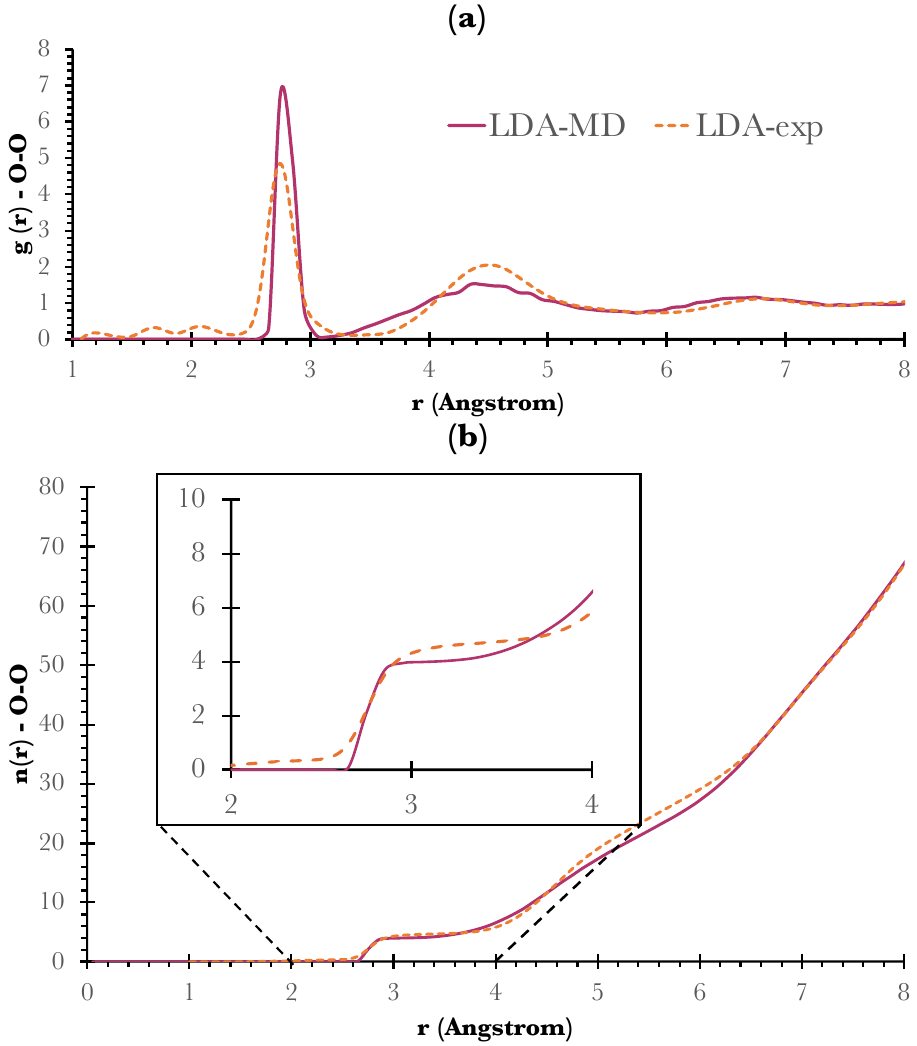}
    \caption{ Comparison between the structure modeled through MD simulation in this work (continuous line), and the experimental results from \cite{Mariedahl} (dashed line), with (a) RDF for the O-O distance and (b) running coordination number for ASW-LDA ices analogues. }
    \label{comp_exp_VS_models_nOO_gOO}
    \end{figure}

The modeled structure has been firstly compared to experimental structural data through the RDF of the O-O distance ($g_{O-O}(r)$), as well as its integrated form, namely the running coordination number, noted $n_{O-O}(r)$. Results are shown in Fig. \ref{comp_exp_VS_models_nOO_gOO}). Empirical data used as comparison are taken from \cite{Mariedahl} who performed X-ray scattering to analyze O-O pair-distribution functions. Let us note that the original data for $g_{O-O}(r)$ were provided in supporting information, but not $n_{O-O}(r)$. The latter has therefore been re-computed according to its definition given in Eq.\,\ref{eq:n(r)}. 

    \begin{equation}\label{eq:n(r)}
          n_{O-O}(r') = 4 \pi \rho \int_{0}^{r'} r^2 g_{O-O}(r) dr
    \end{equation}
        
    \noindent where $\rho$ is the number density of the studied sample. 

Focusing on $g_{O-O}(r)$ (Fig. \ref{comp_exp_VS_models_nOO_gOO}a), the modeled ice aligns well with the experimental curve, providing an initial indication of the suitability of the ice model building approach followed in this work. More specifically, the narrower peaks observed in the modeled RDF, accompanied by higher intensities compared to the experimental data, suggest that the potential used in simulations introduces a slight over-structuration in the modeled ices as compared to empirical protocols. The intrinsic intramolecular rigidity of the TIP4P/2005 force field, which fixes molecular geometry and excludes any intramolecular anharmonic/harmonic oscillations, contributes to the over-structuring effect. However, rigidity is not the sole cause. Flexible force fields also struggle to reproduce experimental \( g_{O-O}(r) \) profiles, as noted in \cite{Ganeshan}. Furthermore, as suggested in \cite{Michoulier}, the over-structuring effect may also arguably stem from the classical treatment of nuclear motion in the simulation methodology, which neglects Zero-Point Energy (ZPE) effects. This is expected to be particularly significant in low-temperature simulations, where ZPE contributions approach/outweigh the available thermal energy, especially for systems involving light elements. Based on the fact that vibrational modes of water molecules display zero-point energies well above room temperature, \cite{Ganeshan} demonstrated that including ZPE contributions in flexible water force fields effectively broadens RDF peaks, taking therefore lower intensities. Omitting any intramolecular flexibility and ZPE effects is therefore expected to lead to under-flexible H-bond network, hence contributing to the observed over-structuring. 

Focusing on $n_{O-O}(r)$ (Fig. \ref{comp_exp_VS_models_nOO_gOO}b), a general concordance is found between our model and the empirical values in \cite{Mariedahl}. In addition, focusing on the [2-4] $\AA$ range, one directly notices the clear plateau for a running coordination number of $\sim$ four, in both the experimental and the modeled curves. This is attributed to the good tetrahedrality of the H-bond network, as deeply discussed in \cite{Michoulier}. 

In the same line, a broader comparison to structural data is presented in Appendix \ref{Appendix_deep_comparison_structural_ice_features}, addressing all pairs' distribution functions with previous MD simulations \citep{Michoulier} and experimental results \citep{Narten_HDA_density,Finney,Mariedahl}, as well as complementary H-bond analyses. These results are fully consistent with the above discussions, and align with theoretical considerations. This validates the suitability of the use of the TIP4P/2005 Force Field in this study.

\subsection{ONIOM-2 scheme}

The choice of the hybrid ONIOM (Our own N-layered Integrated molecular Orbital and Molecular mechanics) scheme (see \citealt{ONIOM_original_paper,ONIOM_gaussian98_implementation,ONIOM_large_syst,ONIOM_QM/MM} and the review from \cite{ONIOM_review} \& references therein) is justified by the large size of the system of interest. Indeed, accurate modeling of binding interactions on amorphous solid water (ASW) requires proper treatment of both long-range and non-local effects, especially for hydrogen-bond cooperativity as discussed in e.g. \cite{Ferrero}. The lack of long-range structural order in ASW, as opposed to crystalline ice, is expected to significantly impact adsorption energetics. To explicitly capture these contributions, the system must therefore be described as extensively as computationally affordable. The use of purely QM methods being consequently proscribed by a too high computational resource requirement, a hybrid/multiscale method has been chosen to handle calculations within a reasonable timescale by splitting the system into multiple fragments. These fragments are then treated using different theoretical methods with distinct levels of theory. 

More specifically, an ONIOM-2 approach has been adopted, proven to be suitable in similar study contexts \citep{Duflot,Tinacci}; the system is divided into 2 layers, (i) the model system and (ii) the real system, representing the whole system. The ONIOM-2 energy is inferred from Eq.\,\ref{eq:ONIOM_Extrapolated_Energy}. 
    \begin{equation}\label{eq:ONIOM_Extrapolated_Energy}
          E_{ONIOM-2 (High:Low)} = E_{(model,High)} + E_{(real,Low)} - E_{(model,Low)}
    \end{equation}
        
\noindent with $E_{ONIOM-2 (High:Low)}$ being the extrapolated ONIOM energy, $E_{(real,Low)}$ quantifying the energy of the real system computed at the low-level of theory and $E_{(model,High)}$ \& $E_{(model,Low)}$ corresponding to the energy of the model system inferred, respectively, from the chosen high and low-level method. 

In the context of this study, ONIOM computations are applied on hemispherical cuts from the MD modeled ice box. Note that these cuts are applied through a custom python script, ensuring that all water molecules included in a given hemisphere are complete, without cut bond. Using the Gaussian software v. 16 \cite[]{Gaussian16}, the actual ONIOM scheme consists in an ONIOM-2(B3LYP-D3(BJ)/6-311+G(d,p):GFN2-xtb external program), as justified in the following. 

The expression used for the computation of binding energies is given by Eq.\,\ref{eq:BE}. Note that the binding energies are corrected for $\Delta$ZPE from harmonic-oscillator frequency computations, and Basis Set Superposition Error (BSSE). The latter accounts for a spurious contribution that comes up in the calculation of interaction energies, leading to an overestimation of the inferred binding energies or, in other words, to an over-stabilization of the complex surface-adsorbate. This arises from the use of finite basis sets, and the superposition of the basis set functions of the interacting fragments (here, (i) the adsorbate, and (ii) the ice system, especially around the binding point), effectively increasing the basis set of each component in the complex. 
    \begin{align}\label{eq:BE}
          BE & = - \Delta E_{Int} \\
          & = - [E_{complex} -  E_{ice} - E_{ads} + \Delta E_{CP}^{BSSE *} + \Delta ZPE^{*}]  \notag \\
          & = BE_{SCF} + \Delta E_{CP}^{BSSE} + \Delta ZPE \notag \\
          \text{with\hspace{0.15cm}} & \Delta ZPE^{*} = (E_{ZPE, complex} - E_{ZPE,ice} - E_{ZPE,ads}) . f_{scaling}  \notag \\
          & = - \Delta ZPE \notag
        \end{align} 
    
\noindent where $\Delta E_{Int}$ is the interaction energy, $E_{complex}$ is the energy of the complex adsorbate-ice, extrapolated from ONIOM computations (Eq.\,\ref{eq:ONIOM_Extrapolated_Energy}), $E_{ice}$ is the ice energy, also extrapolated from ONIOM computations, $E_{ads}$ is the energy of the adsorbate, computed at the high-level of theory used for the ONIOM scheme. $\Delta E_{CP}^{BSSE (*)}$ is the counterpoise-correction by Boys and Bernardi for the BSSE computed on the model zone; BSSE is actually expected to be non-zero only near the adsorption site. Indeed, regarding the finite size of the used basis set, as the adsorbate approaches the substrate, its basis functions are susceptible to overlap those from water molecule atoms from the surface. This overlap is therefore most likely to occur around the binding site. Note that $\Delta E_{CP}^{BSSE *}$ and $\Delta E_{CP}^{BSSE}$ have opposite signs, typically taking positive and negative values, respectively. $E_{ZPE,i}^{(*)}$ is the ZPE energy of fragment i (computed under the harmonic oscillator hypothesis), multiplied by $f_{scaling}$, i.e. the frequency scaling factor. This factor depends on both the employed computational method (here, the DFT functional) and the basis set, and is taken from the NIST database\footnote{\url{https://cccbdb.nist.gov/vibscalejust.asp}}. $BE_{SCF}$ stands for the BE values without taking into account the BSSE and ZPE contributions.

\subsubsection{Model zone setup}

\paragraph{(a) Model zone size.} The high-level (HL) size $R_{Model, HL}$ has already been constrained in \cite{Tinacci} in similar systems, in which a benchmark on the model zone size on a cluster of 200 water molecules has been performed. Note that in contrast with the ASW model building method in the present study, the cluster in \cite{Tinacci} is made through a bottom-up approach, without PBC. In this paper, the BE computation is made onto hemispheric sub-water clusters extracted from the larger ASW analogue, taking advantage of the PBC along x and y axis. This benchmark scanned $R_{Model, HL}$ from 5 to 8.5 $\AA$ on 8 different optimized binding sites. Results showed that BE slowly evolves with increasing model zone size, showing different evolution tendencies over the studied binding sites, with variations within 5 kJ/mol, or slightly above. In this scope, in order to save computational resources, the smallest $R_{Model, HL}$,  i.e. 5 \AA, corresponding to the adsorbate plus 3 to 9 H$_2$O depending on the binding sites has been chosen. However, upon closer examination of the results from \cite{Tinacci}, the BE convergence w.r.t. the model zone size is located around 7-8 $\AA$, corresponding to 11 to 36 H$_2$O depending on the binding site. Moreover, with regard to the small model zone within the method proposed by \citet{Tinacci}, system rearrangements forced iterative redefinitions of the water molecules included in the model zone. Consequently, in the scope of this study, $R_{Model, HL}$ is set as a starting point to 8 \AA, corresponding to 20-25 water molecules (plus the adsorbate). In other words, all H$_2$O lying within 8 $\AA$ from the adsorbate barycenter, without any cut bond, are included in this model zone. This choice is further supported by the study by \cite{Duflot}, who performed BE computations on hemispheric cuts, with $R_{Model, HL}$ equal to 7-8 \AA, including around $\sim$ 20 water molecules. Moreover, in \cite{Song_Kastner,Song_Kastner_2017}, the model zone includes all water molecules with at least one atom within 6 \AA, i.e. 23 H$_2$O, also using a hemispheric cut. This is therefore analogous to the model zone size setup in this work. 

It is worth mentioning that, to get a converged result, the choice of $R_{Model, HL}$ depends on the system, including the structure of the ice model, the adsorbate, and the sampled binding site, but also on the real system size and the low-level method, as well as the high-level method. For that reason, the value of 8 $\AA$ for the model zone size will be retro-checked downstream the benchmark on the low-level size, discussed in Sect. \ref{Low_Lev_Design_Sect}.

\paragraph{(b) High-level QM method.}
The model zone is treated under a DFT scheme. More specifically, the hybrid exchange-correlation B3LYP functional \citep{Lee} along with the D3 version of Grimme’s dispersion correction with the Becke-Johnson damping function \citep{GrimmeDcorr,GrimmeBJ} has been used (D3(BJ) in the following), as in \cite{Ferrero} where the BEs of a set of relevant interstellar species have been computed through periodic calculations. This choice of functional was based on results in \cite{Kraus&Franck} in which B3LYP-D3(BJ) has been shown to provide a good level of accuracy for the interaction energy of non-covalently bound dimers, with a mean absolute error of 0.06 $\AA$ w.r.t. semi-empirical the bond length data. 

    \begin{table}[]
        \centering
        \caption{DFT functional benchmark results with respect to CCSD(T) over the three adsorbate test cases.}
        \begin{tabular}{l l l }
        \hline \hline
            DFT Functional & MARD (\%) & MARD (\%) \\
            & 6-311+G(d,p) & aug-cc-pVTZ \\ \hline
            B3LYP-D3(BJ) & 13.4 & 7.1 \\
            PBE0-D3(BJ) & 51.3 & 12.3 \\
            PBE0 & 22.8 & 30.1 \\
            BMK-B3(BJ)  & 145.0 & 93.6 \\
            BMK & 89.0 &  83.8 \\
            B3PW91-D3(BJ) & 24.9 & 10.4 \\
            B3PW91 & 84.9 & 89.8 \\
            CAM-B3LYP & 18.6 & 33.0 \\
            M06-2X & 22.7 & 9.6 \\
            B97-D & 22.8 & 15.2 \\
            B97-D3(BJ) & 24.2 &  7.7 \\
            $\omega$B97X-D & 17.8 & 12.6 \\ \hline
        \end{tabular}
        \label{tab:HL_Functional_benchmark_summary_results}
    \end{table}

The use of this functional has been further validated by means of a benchmark on tetrameric binding configurations, implying 3 water molecules plus a given adsorbate. Two basis sets are compared, namely 6-311+G(d,p) and aug-cc-pVTZ. In practice, for each studied adsorbate, one tetrameric structure per species is extracted from optimized adsorbate-ice complexes and its isolated counterparts. Subsequent single-point energy evaluations have been performed to infer the interaction energies, including the BSSE contribution (but not the $\Delta$ZPE); actually, the BSSE amplitude depends on the chosen basis set, and this dependency itself depends on the level of theory. More specifically, wavefunction-based CCSD(T) is more sensitive to the finite size of the basis set than DFT methods. Using CCSD(T) as reference, results from 8 DFT functionals (B3LYP-D3(BJ); PBE1PBE-D3(BJ); PBE0; BMK-D3(BJ); BMK, B3PW91-D3(BJ); B3PW91; CAM-B3LYP; M06-2X; B97-D3(BJ); B97-D\footnote{The native implementation of B97 functional in Gaussian is defined along with the D2 dispersion from the Grimme's group.}; $\omega$B97X-D) have been compared with or without dispersion correction depending on their respective parameterization. The performance of each functional relative to the CCSD(T) results varies depending on the studied interaction type and the adsorbate. The main objective pursued in this study is the development of a method applicable to several relevant closed shell interstellar species. Therefore, this benchmark approach has the aim to select a functional that provides an appreciable estimate over different molecular interactions. In this regard, the prime interest is directed towards the Mean Absolute Relative Difference (MARD) over the three adsorbates with respect to the CCSD(T) value for each functional, as presented in Table \ref{tab:HL_Functional_benchmark_summary_results}. Quantitative details are given in Appendix \ref{Appendix_HL_functionnal_details}. From these results, B3LYP-D3(BJ) provides the best estimations for both basis sets. The relative differences between the 6-311+G(d,p) series and the improved aug-cc-pVTZ performances arise primarily from the BSSE correction contribution applied to the reference value. For instance, with the 6-311+G(d,p) basis, in the case of CO, the BSSE CSSD(T) correction (4.2 kJ/mol) approaches the same magnitude as the interaction energy itself (5.6 kJ/mol), underlining the importance of carefully considering BSSE and its own uncertainty when interpreting benchmark results for such systems. Considering the more extended aug-cc-pVTZ basis set, this contribution reduces to 2.2 kJ/mol, for BE value of 8.5 kJ/mol. Given the ultimate goal of extensively sampling an amorphous water surface with relatively large water sub-clusters, B3LYP-D3(BJ)/6-311+G(d,p) as high-level of theory nevertheless offers a good compromise for an appropriate accuracy/computational cost balance. A final argument in favor of this choice, i.e. the convergence between the B3LYP-D3(BJ)/6-311+G(d,p) and B3LYP-D3(BJ)/aug-cc-pVTZ results, is discussed in Appendix \ref{Appendix_high_level_BS_against_growing_LowLevel}.

    \begin{table*}[]
        \centering
        \caption{\label{tab:contrib_BSSE_ZPE_benchmark_low_level_size}BEs for growing $\Delta$R$_{LL}$ and fixed R$_{Model, HL}$ (8 $\AA$)  at ONIOM(B3LYP-D3(BJ)/6-311+G(d,p):xtb) level, along with the respective contributions from BSSE and $\Delta$ZPE and BE$_{SCF}$  (see Eq. \ref{eq:BE}).}
        \begin{tabular}{lllllllll}
        \hline \hline
         Adsorbate &
          \begin{tabular}[c]{@{}l@{}}Contributions\\ (kJ/mol)\end{tabular} &
          $\Delta$R$_{LL}$ 3 Å &
          $\Delta$R$_{LL}$ 4 Å &
          $\Delta$R$_{LL}$ 5 Å &
          $\Delta$R$_{LL}$ 6 Å &
          $\Delta$R$_{LL}$ 8 Å &
          $\Delta$R$_{LL}$ 10 Å &
          $\Delta$R$_{LL}$ 11 Å \\
          \hline
        \multirow{3}{*}{\begin{tabular}[c]{@{}l@{}}NH$_3$ – 1st\\ Binding site\end{tabular}} &
            BE &
          49.8 &
          51.2 &
          50.1 &
          51.4 &
          53.2 &
          53.0 &
          53.1 \\
         & BE$_{SCF}$ &
           68.1 &
           70.3 &
           68.9 &
           70.3 &
           72.0 &
           71.8 &
           71.9 \\ 
         & $\Delta E_{CP}^{BSSE}$  & - 4.9           & - 4.9  & - 4.8  & - 4.8  & - 4.9  & - 4.9  & - 4.9  \\
         & $\Delta$ZPE & - 13.4          & - 14.2 & - 14.0 & - 14.1 & - 13.9 & - 13.9 & - 13.9 \\
         \hline
        \multirow{3}{*}{\begin{tabular}[c]{@{}l@{}}CO – 1st\\ Binding site\end{tabular}} &
           BE &
          8.5 &
          11.2 &
          11.4 &
          11.2 &
          10.8 &
          11.2 &
          11.2 \\
         & BE$_{SCF}$ &
           12.5 &
           15.2 &
           15.5 &
           15.2 &
           14.9 &
           15.3 &
           15.3 \\
         & $\Delta E_{CP}^{BSSE}$  & - 2.0           & - 1.7  & - 1.8  & - 1.8  & - 1.8  & - 1.8  & - 1.8  \\
         & $\Delta$ZPE & - 2.0           & - 2.3  & - 2.3  & - 2.2  & - 2.3  & - 2.3  & - 2.3  \\
         \hline
        \multirow{3}{*}{\begin{tabular}[c]{@{}l@{}}CH$_4$ – 1st\\ Binding site\end{tabular}} &
           BE &
          8.4 &
          8.3 &
          8.6 &
          9.1 &
          9.8 &
          10.3 &
          10.5 \\
         & BE$_{SCF}$ &
           13.7 &
           14.0 &
           14.3 &
           14.7 &
           15.5 &
           15.8 &
           16.0 \\
         & $\Delta E_{CP}^{BSSE}$  & - 1.1           & - 1.1  & - 1.1  & - 1.1  & - 1.2  & - 1.2  & - 1.2  \\
         & $\Delta$ZPE & - 4.2           & - 4.6  & - 4.6  & - 4.5  & - 4.5  & - 4.3  & - 4.3  \\
         \hline
        \multirow{3}{*}{\begin{tabular}[c]{@{}l@{}}NH$_3$ – 2nd\\ Binding site\end{tabular}} &
          BE &
          80.7 &
          20.8 &
          21.1 &
          22.4 &
          22.2 &
          23.1 &
          23.4 \\
         & BE$_{SCF}$ &
           99.3 &
           28.9 &
           29.9 &
           31.0 &
           31.0 &
           31.9 &
           32.1 \\
         & $\Delta E_{CP}^{BSSE}$  & - 5.5  & - 4.2  & - 4.2  & - 4.1  & - 4.1  & - 4.1  & - 4.1  \\
         & $\Delta$ZPE & - 13.1 & - 3.9  & - 4.6  & - 4.5  & - 4.7  & - 4.7  & - 4.6 \\\hline
        \end{tabular}
    
    \end{table*}

\subsubsection{Low-level design}\label{Low_Lev_Design_Sect}

\paragraph{(a) Real system size benchmark.}
So far, very few studies have addressed the impact of the system size, or more specifically the value of the real system size, on the inferred BEs. However, an appropriate choice of the model zone size $R_{Model, HL}$ and of the total system size ($R_{Hem, Real} = R_{Model, HL} + \Delta R_{LL}$, LL standing for the low-level of theory) are also of prime importance to obtain convergent BEs. Indeed, as previously stated, to accurately model binding interactions on ASW systems, one should properly take into account the contributions from long-range and non-local interactions, with a prime emphasis on H-bond cooperativity \citep{Ferrero}. In order to properly include such contributions, the real system size has consequently been benchmarked, with low-level shell size $\Delta R_{LL}$ ranging between 3 and 11 \AA, resulting in real system size $R_{Hem,Real}$ expanding from 11 to 19 \AA. Note that, throughout this study, the molecules in the $\Delta R_{LL}$ shell are kept frozen. This freezing of the low-level zone has also been applied by \citet{Duflot} and \citet{Tinacci}. This artificially preserves the ice stiffness experienced in real size systems far from the binding zone while reducing the computational cost compared to a flexible low-level zone.   

The structure of each of these systems is optimized through ONIOM computations, with the semiempirical GFN2-xtb algorithm from the Grimme's group \citep{GFN2-xtb,xtb} as low-level method. These computations are therefore performed at ONIOM(B3LYP-D3/6-311+G(d,p) : GFN2-xtb) level. $\Delta$ZPE and BSSE contributions are included. For the model zone, Gaussian16 tight Self-Consistent-Field (SCF) convergence criteria are used. For the low-level treatment, as xtb is not implemented in Gaussian \cite[]{Gaussian16}, it has been called through the standardized wrapper interface \citep{Beaujean_pemices_wrapper_g16} for communication with external program proposed by Gaussian16, with default SCF parameters. The three adsorbate test cases have been studied at a first binding site, and another binding site has also been studied with NH$_3$. In fact, NH$_3$ is known to present a large BE dispersion compared to CO and CH$_4$. Studying a second binding site with distinct binding behavior is therefore relevant to extrapolate the conclusions for the final sampling of the BE distributions. The results are given in Table \ref{tab:contrib_BSSE_ZPE_benchmark_low_level_size}. 

For NH$_3$ on a first site, the BEs are converging from $\Delta R_{LL}$ of 8 \AA, corresponding to a total radius of 16 \AA, at a value of $\sim$ 53 kJ/mol. At $\Delta R_{LL}$ equal to 6 \AA, one underestimates the converged results only by less than 2 kJ/mol, which accounts for less than 4\% of the total binding energy for this case. Regarding CO, the convergence already appears at $\Delta R_{LL}$ of 4 $\AA$. On the other hand, CH$_4$ BE values show a slower convergence, appearing only from a $\Delta R_{LL}$ of 10 \AA, at a BE of $\sim$ 10-3-10.5 kJ/mol. A $\Delta R_{LL}$ of 8 $\AA$ leads to an under-estimation of 0.5-07 kJ/mol (< 7\%) with respect to that converged value, while a $\Delta R_{LL}$ of 6 $\AA$ underestimates the converged value by 1.2-1,4 kJ/mol ($\sim$ 13\%). In the case of NH$_3$ at a second binding site, the convergence of the BE value takes place at $\Delta R_{LL}$ of 10 \AA, at $\sim$ 23-23,5 kJ/mol. With $\Delta R_{LL}$ of 6 to 8 \AA, one underestimates this converged value only by $\sim$ 1 kJ/mol, representing $\sim$ 5\% of the total value of NH$_3$ BE at this site. Focusing then on the smallest system size, with a $\Delta R_{LL}$ of 3 \AA, the computed binding energy drastically increases toward a value of more than 80 kJ/mol. It can be arguably rationalized by geometrical artifacts due to the asymmetry of the extracted hemisphere caused by its small size. The system asymmetry rapidly decreases with increasing system size. Furthermore, one may also suggest a potential too small effective coherence/rigidity of the ice system at such a small system size. Indeed, as the low-level zone (frozen) is embedding the model zone, a too small $\Delta R_{LL}$ may in some cases not be representative of the embedding environment from a real ice system and its intrinsic H-bond cooperativity, enhancing its coherence.

Taking all these considerations into account, $\Delta R_{LL}$ of 8 $\AA$ seems to be the best compromise between accuracy and computational cost. It corresponds to $\sim$ 210-225 H$_2$O molecules in the $\Delta R_{LL}$ frozen ice shell, and therefore a total number of 235-250 water molecules are included in the entire ice hemisphere. This is similar to the system size of \cite{Tinacci}, in which they used a water cluster of 200 molecules. 

Regarding finally the contributions of the applied correction, $\Delta$ZPE generally accounts for $\sim$ 20-25 \% of BE, and can go up to 50\% for weakly bound systems as in the case of CH$_4$ (apolar). On the BSSE side, their contribution to the inferred BEs is more variable, ranging from 5-10 to 25\% depending on the adsorbate and the binding site. Note also that both of these contributions are very stable w.r.t. the $\Delta R_{LL}$ value.

The same benchmarking analyses have been conducted with 3 other basis sets for the model zone description, i.e. def2-TVZPP, aug-cc-pVDZ and aug-cc-pVTZ. This has the aim of quantifying the stability of the computed BEs with growing real system size against different basis set for the high-level. The results are discussed in Appendix \ref{Appendix_high_level_BS_against_growing_LowLevel}. From a general perspective, across the 3 adsorbate test cases, there is a consistently good match of the results obtained with ONIOM(B3LYP-D3(BJ)/6-311+G(d,p):xtb) and ONIOM(B3LYP-D3(BJ)/aug-cc-pVDZ:xtb) or ONIOM(B3LYP-D3(BJ)/aug-cc-pVTZ:xtb). This indicates that 6-311+G(d,p) is suitable to capture the essential features of the electronic structure relevant to this study. 

\paragraph{(b) GFN2-xtb performance for the low-level description.}
Although the convergence of the BEs with respect to the real system size has been investigated, it is nevertheless worth to question the reliability of the employed low-level method, i.e. the semi-empirical GFN2-xTB quantum mechanical algorithm developed by the Grimme group \cite[]{GFN2-xtb,xtb}. The general performances of GFN2-xtb have already been widely investigated and validated, such as in the original \cite{GFN2-xtb} paper. It has also been evaluated in \cite{Germain} and \cite{Tinacci} on similar amorphous water systems than in our context of study. In order to gain additional method-specific insights into the stability of xtb performance against growing $\Delta R_{LL}$, ONIOM2(DFT:xtb) BE results are compared to ONIOM2(DFT:DFT) results with growing real system sizes, as deeply commented in \ref{Appendix_xtb_performances_check}. While theses results showed an overall good stability of the xtb performances w.r.t. to DFT treatments over growing real system size at BE-real system size convergence, they also provides further justifications for the previously commented choice of the $\Delta R_{LL}$ size. Theses are complementary to the analysis from \cite{Germain} and \cite{Tinacci}.

\subsubsection{Retro-verification of the ONIOM setup (layer sizes)}

In order to retro-check the chosen model and real system sizes, taking advantage of the 2D (x-y) periodic boundary conditions of the final box from the MD simulations, computations of binding energies are performed on even larger hemispheric cuts, with $R_{Hem,Real}$ of 25 $\AA$ for a model zone size of 12 $\AA$. BE$_{SCF}$ has been recomputed for each of the three adsorbate test cases at the first sampled binding site, in order to compare the results with the converged BE$_{SCF}$ values at largest benchmarked real system size, as reported in Table \ref{tab:contrib_BSSE_ZPE_benchmark_low_level_size}. Indeed, this retro-checking is highly resource-consuming; knowing that $\Delta$ZPE and BSSE corrections are stable against the system size, these have not been considered in this part of the study.  

For NH$_3$, BE$_{SCF}$ amounts to 72.8 kJ/mol, only differing by $\sim$ 0.9 kJ/mol (less than 1.5 \%) as compared to the value of 71.9 kJ/mol for the largest studied system size in Table \ref{tab:contrib_BSSE_ZPE_benchmark_low_level_size}. In the case of CH$_4$, a value of 16.9 is obtained from this retro-checking, compared to 16.0 kJ/mol. Therefore, the relative difference accounts for less than 6 \%. Concerning CO, a value of 15.6 kJ/mol is obtained, in complete agreement with the rapid convergence of the CO binding energies with respect to the cluster size at a BE of 15.3 kJ/mol. These results validate the choice of high- and low-level sizes for the ONIOM scheme design.

\subsection{Twofold BE sampling} 
The sampling of the binding sites and starting configurations has been performed through a custom python script, randomizing the substrate exploration in terms of binding sites and adsorbate-to-substrate orientations. More specifically, the underlying sampling methodology is twofold, distinguishing two main sources of BE diversity, i.e. (i) the surface typology and its complexity, itself related to the amorphicity of the underneath ice, and (ii) the substrate-to-adsorbate orientations, and the associated local roughness of the potential energy landscape at a given binding site. The first source can be sampled unbiasedly by placing a regularly spaced grid above the surface to define the starting position of the adsorbate barycenter for subsequent geometrical optimizations. The second source of binding diversity is sampled by considering various adsorbate orientations for each spatially distinct binding site. After geometry optimization, the computed BEs can in fact vary due to distinct optimized configurations\footnote{The optimized configuration is defined by both the final adsorbate-to-substrate orientation and the associated feedback onto the relative orientation of the surrounding water molecules after geometry optimization.}, thereby contributing to the local roughness of the potential energy surface (PES) at each sampling grid point. In other words, a binding zone associated to a given sampling grid point is not a perfectly smooth well within the substrate PES, but rather displays a kind of sub-energy-landscape accounting for different adsorbate-to-substrate optimized configuration. The consideration of this second source of heterogeneity to the energy landscape of an amorphous surface is important to comprehensively sample its diversity. Note that, in the method proposed in \cite{Tinacci}, the adsorbate orientation is randomized, but considering a single orientation per site. 

In practice, this custom python script automates the workflow from the hemispheric cuts in the MD-modeled water box, to the ONIOM-2 input writing within the ONIOM scheme designed upstream the BE sampling, their submission and monitoring. More details on the steps followed by this script are provided in Appendix \ref{Appendix_automated_sampling_script}. Note that, for each binding site, the number of adsorbate-to-substrate starting orientation(s) per adsorbate depends on the adsorbate symmetry; for NH$_3$, 3 starting orientations have been studied, with (i) the three hydrogens facing the surface, (ii) the 3 hydrogens flipped in the opposite direction in z, and (iii) the plane formed by the three hydrogens being perpendicular to the surface. In the case of CO, 3 orientations have also been sampled, with (i) CO axis parallel to the surface, (ii) CO axis parallel to the normal to the surface, with O atom facing the surface, and (iii) the opposite. Finally, because of its higher order symmetry, only one orientation has been studied for CH$_4$.

The binding configuration redundancy reduction deserves further comments. On the one hand, concerning juxtaposed binding sites, the separation distance between points in the sampling grid has been chosen large enough to avoid any redundancy between adjacent sites, minimizing the probability that the NH$_3$ position evolves towards the same energy minimum (same implied water molecules from the surface) as the juxtaposed system during the geometry optimization. A RMSD analysis confirming this point has been performed, as discussed in Appendix \ref{Appendix_RMSD_analysis}. On the other hand, regarding the case of different starting orientations of the adsorbate on a given hemispheric cut converging toward the same energy minimum, this redundancy has been reduced to the distinct binding configurations per binding site. More precisely, such an in-site redundancy has been identified based on the $\Delta$BE and RMSD values between pairs of starting adsorbate-to-substrate orientations, i.e. defining $\Delta$BE$_{ij}$ < 0.05.<BE>$_{ij}$ kJ/mol and RMSD < 1 $\AA$ as redundancy cutoff, where <BE>$_{ij}$ is the mean BE for i and j orientations. Further comments are given in Appendix \ref{Appendix_RMSD_analysis}. In this idea, the resulting distribution refers to the sampling of each unique trap within the potential energy landscape of the substrate, both in terms of locally distinct sites and local traps.

\section{Results - Newly inferred distributions versus previously reported values}
\label{sec:Results}

\subsection{NH$_3$ as adsorbate}

After geometry optimization and harmonic frequency computation, only 9 geometry optimizations of adsorbate/ice systems out of 300 did not converge under the defined cutoffs, and 25 systems showed imaginary frequencies, of which 14 showed 1 to 3 imaginary frequencies in the range $i1$ to $i27$ cm$^{-1}$. Because of the small values of these imaginary frequencies, they are not associated with NH$_3$ nuclear motions and do not contribute significantly to the calculation of ZPEs, as justified by \cite{Tinacci}. Based on these considerations, these 14 systems are also retained. This yielded 280 zero-point and BSSE-corrected BEs for NH$_3$ interactions with the modeled ASW ice analogue. After reduction for in-site binding configuration redundancy, the final data set consists of 204 BE values.  

\begin{figure}[]
    \centering
    \includegraphics[width=0.9\linewidth]{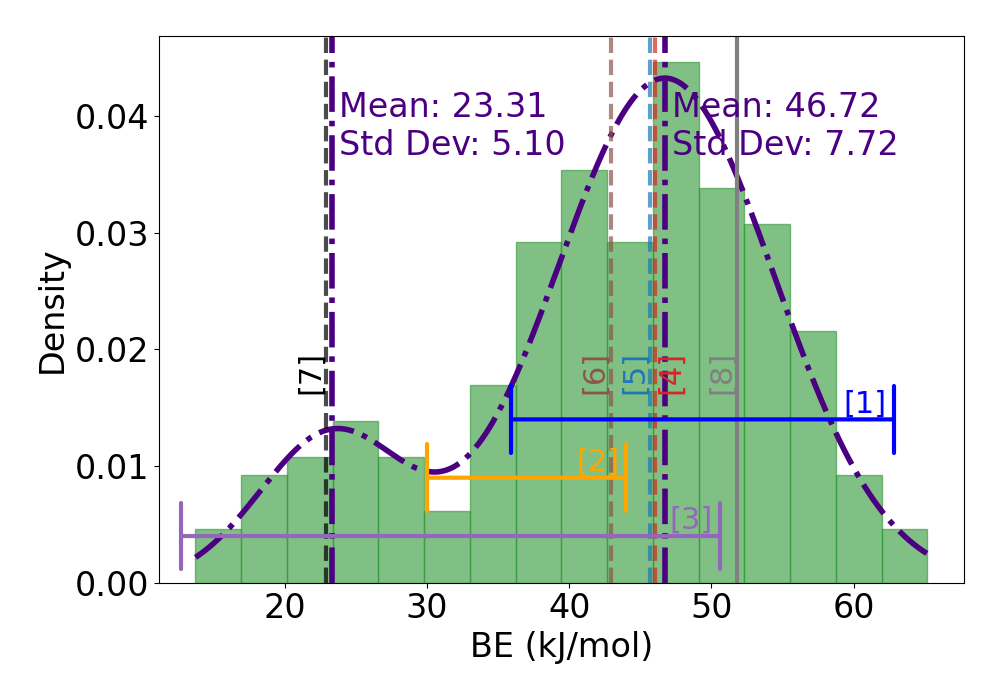} 
    \caption{\label{NH3_distr_corr_BEs_hist}Normalized histogram of sampled NH$_3$ BE (in kJ/mol), fitted with a Gaussian Mixture Model using scikit-learn algorithm \cite[]{scikit-learn}. Previously reported NH$_3$ BE values onto water ice analogs with diverse simulation methods are given on top. Horizontal intervals report on published BE dispersions, while vertical dashed lines relate to works focusing on single BE values. Methods used in each work are discussed in the text. The BE value on crystalline ice from \cite{Ferrero} is also given (plain gray line). Ref: [1] \cite{Ferrero}; [2] \cite{Duflot}; [3] \cite{Tinacci}; [4] UMIST , [5] \cite{Wakelam}/KIDA; [6] \cite{Das}; [7] \cite{Penteado}; [8] \cite{Ferrero} on crystalline ice}
    \end{figure}

The resulting normalized distribution is given in Fig. \ref{NH3_distr_corr_BEs_hist}, on top of which a Gaussian Mixture Model (GMM, in mauve) is fitted. Indeed, standard Gaussian fits have been tested as well as other non-symmetric fit types, but it turns out that the data are much better described by this GMM profile. Previously reported single NH$_3$ BE values (dashed/plain vertical lines) or dispersion (horizontal intervals) are also reported.  

In terms of distributions, Fig. \ref{NH3_distr_corr_BEs_hist} reports the dispersion from \cite{Ferrero} who studied the binding behavior of interstellar species on an ASW model through PBC B3LYP-D3(BJ)//HF-3c composite computations, i.e. with geometry optimization with HF-3c method \citep{HF-3c}, and subsequent single point energy evaluation at B3LYP-D3(BJ) level of theory. The resulting NH$_3$ BE distribution ranges from 35.9 to 62.8 kJ/mol. Note that from the amorphous model in \cite{Ferrero}, made of a unit cell with 60 H$_2$O, only 7 binding sites have been studied. Moreover, the starting positions are chosen based on electrostatic potential map to maximize the chance of a strong H-bond interactions. As compared to this work, results reported in \cite{Ferrero} only cover the range characteristic of the main peak from the GMM fit in Fig. \ref{NH3_distr_corr_BEs_hist}. This arguably comes from the lower number of sampled binding sites while biasing the adsorbate position by optimizing the adsorbate surface orientation. All NH$_3$ binding configurations in \cite{Ferrero} presented NH$_3$ as acceptor of a H-bond, explaining that no contribution from the lower energy part of the NH$_3$ BE distribution presented here is included in the results reported in \cite{Ferrero}. The BE dispersion from \cite{Duflot} is also given on Fig. \ref{NH3_distr_corr_BEs_hist}, where ONIOM(CBS/DLPNO-CCSD(T):PM6)//ONIOM($\omega$B97X-D/6-31+G**:PM6) ZPE-corrected BE computations (no BSSE correction) are performed on hemispheric cuts with a model and real system sizes of 8 and 12 to 15 \text{\AA}, respectively. This gives from 140 to 170 H$_2$O included in the hemisphere. The initial positions for a given adsorbate are chosen from upstream classical molecular dynamics trajectories at 77K, modeling the adsorption of single water molecules with a random starting position of the adsorbed water molecules for each trajectory. As compared to the sampling method is the present paper, it intrinsically hinges on the water molecules adsorption behavior at such temperature conditions. This therefore accounts for a more biased sampling of the binding site since water molecules are more likely to reorient and, to a lesser extent diffuse, at 77K. As compared to the distribution from this work, results in \cite{Duflot} cover a much narrower range of energy, with BE values of 35.8 $\pm$ 3.4 kJ/mol. The work by \cite{Tinacci} has also been considered, a paper extensively discussed upstream due to its similarities with the method adopted in the present paper. Indeed, the method developed in \cite{Tinacci} aims to sample the NH$_3$ BE distribution by unbiasedly defining the adsorbate initial position, while randomizing its orientation w.r.t. the surface with one random orientation per sampled position. The underneath BE inference method consists of an ONIOM(DLPNO-CCSD(T)//B97D3(BJ):xTB-GFN2) approach, with a model zone of 5 \text{\AA}, i.e. including $\sim$ 3 to 10 water molecules. The ASW model is constituted of an agglomerated cluster of 200 water molecules built from successive MD additions of H$_2$O and stepwise geometry optimizations. After binding site redundancy reduction, the resulting NH$_3$ BE distribution includes 77 values from the 162 starting points. As compared to results in \cite{Ferrero} and \cite{Duflot}, the distribution in \cite{Tinacci} is much wider, leading to a BE values of 31.1 $\pm$ 8.8 kJ/mol. This fully covers the low-energy tail from the results presented in this paper, but does not reproduce the higher-energy tail. Note that the methodology proposed in \cite{Tinacci} implies the sampling of full surface of the 200-H$_2$O icy model by projecting a uniformly spread grid of points. In the present paper, hemispheric clusters have been extracted from a larger water box. This aims to ensure to reach bulk properties below the sampled surface, as well as avoiding local edge effects. Moreover, the model zone in \cite{Tinacci} was smaller than in this work. The deviations between results from these works and the present study are probably explained by the number of sampled binding interactions, combined to the chosen sampling grid spacing, and by the study of 3 adsorbate starting orientations per grid point. 

Furthermore, single value references are also given in Fig. \ref{NH3_distr_corr_BEs_hist}. The plain gray line represents the BE value found in \cite{Ferrero} in which they also studied the binding behavior of interstellar species on a periodic crystalline proton-ordered ice slab model, with a BE value of 51.8 kJ/mol. Although it falls within the BE distribution from this work, it is larger than the mean of the main peak. This trend has also been highlighted in \cite{Ferrero}. This is simply due to the higher rigidity of the crystalline phase, which enforces the typical binding interactions through a smaller distortion energy cost upon adsorption. Dashed lines report on single BE values from the UMIST/KIDA databases, as well as the computational work by \cite{Wakelam}, \cite{Das} and \cite{Penteado} where BE was inferred from the empirical study by \cite{Collings}. It is interesting to highlight the value found in \cite{Penteado}, surprisingly falling in the middle of the lower energy subpeak of the distribution reported here. However, this is computed from inversion curves from TPD experiments and greatly depends on the adopted pre-exponential factor.  

From an overall perspective, this comparison demonstrates that all single values and energy ranges covered by the previously published NH$_3$ BE distributions are encompassed by the NH$_3$ BE dispersion resulting from the present study. In contrast, none of those previously published distributions individually covers the full range spanned by the results reported here. A combination of these distributions is required to achieve this coverage. It thereby allows for a kind of consensus that could not previously be established as a result of the disparity in the range of energetic values previously covered. Combined with the special attention devoted to the statistical convergence of the inferred distribution (see Sect. \ref{sec:Discussions}), compliant with the expectations of improvements from the method presented in this paper, this lends support to the credibility of the newly proposed NH$_3$ BE distribution. Let us nevertheless remind the different levels of theory/method of BE inference among the compared works, and their inherent uncertainty. Moreover, this clearly highlights the significant challenge faced by the astrochemical community regarding the high uncertainty in BEs. The frequent assumption of a single value per adsorbate in chemical networks drastically impacts inferred surface chemistry, as discussed and studied in \cite{Grassi}. In the case of NH$_3$, given the observed large BE distribution dispersion, its consideration should have a substantial effect on the results of the associated surface chemistry. Furthermore, it is known that in molecular clouds, NH$_3$ is a key component of water-rich icy mantles. Therefore, accurately considering its BE distribution, including the first peak at low BEs, is of paramount importance, as further commented in Sect. \ref{sec:astro_implic}.

\subsection{CO as adsorbate}

\begin{figure}[]
    \centering
    \includegraphics[width=0.9\linewidth]{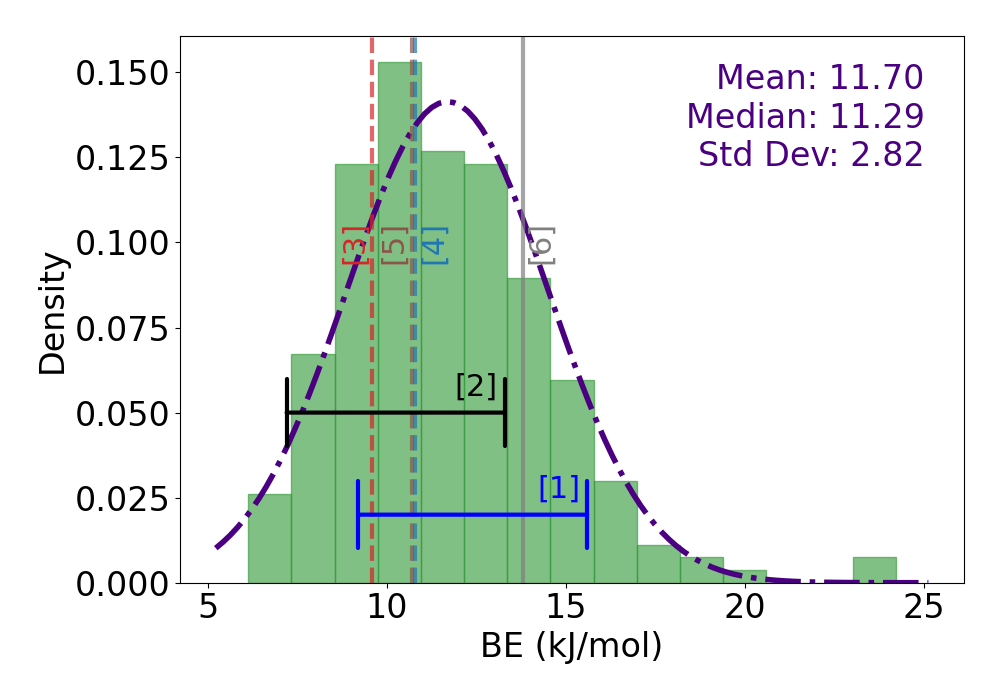} \caption{\label{CO_distr_corr_BEs_hist}Normalized histogram of sampled binding energies (in kJ/mol) for CO, fitted with a single component gaussian profile. Previously reported results of CO BE values onto water ice analogs with diverse simulation methods are given on top. See Fig \ref{NH3_distr_corr_BEs_hist} for the meaning of horizontal intervals, vertical dashed lines and the plain gray line. Ref: [1] \cite{Ferrero}; [2] \cite{He} \& \cite{Penteado}; [3] UMIST; [4] \cite{Wakelam}/KIDA, [5] \cite{Das} ; [6] \cite{Ferrero} on crystalline ice}
    \end{figure}

After geometry optimization and harmonic frequency computation, 19 geometry optimizations of adsorbate/ice systems out of 300 did not converge within the defined cut-offs, and 33 systems exhibited imaginary frequencies, of which 18 configurations showing one to max three imaginary frequencies in the range $i3$ to $i19$ cm$^{-1}$ have been kept. 266 systems have provided valid BE values for the study of CO binding behavior on ASW ices. After reduction for in-site binding configuration redundancy, the final dataset for CO is composed of 228 BE values. The associated normalized distribution is given in Fig. \ref{CO_distr_corr_BEs_hist}, on top of which a Gaussian profile is fitted (mauve). As expected, the mean BE is significantly below the typical values for NH$_3$. This is related to the absence of H-bonding interactions with CO because of its strong internal bonding and the associated electron distribution that prevents it from forming significant H-bonds with water molecules. Previously reported single CO BE values or dispersion are also reported in Fig. \ref{CO_distr_corr_BEs_hist}. 

In terms of distributions, results from \cite{Ferrero}, in which 5 binding sites for CO have been sampled, as well as experimental TPD results from \cite{He} and \cite{Penteado} have been considered. Both of them mostly cover the central part of the distribution reported in this work, but none includes the left and right tails. This is certainly explained by the higher number of sampled binding sites as well as the systematic study of multiple starting adsorbate orientations in the method proposed in the present work. This is supported by the great attention devoted to the converged nature of the inferred distribution, and the associated difficulty to accurately reproduce the slowly converging standard deviation (see Sect. \ref{sec:Discussions}). 

Regarding previously published single values for CO, the crystalline CO BE from \cite{Ferrero} (plain grain line) falls to the mid-high side of our distribution, a trend also emerging from the analysis of the NH$_3$ distribution, as well as from the results in \cite{Ferrero}. Dashed lines concern single BE values from the UMIST/KIDA databases, as well as from the computational work from \cite{Wakelam} and \cite{Das}, all of them falling within the energy range covered in this work, slightly below the computed global mean/median. 

As for NH$_3$, this comparison showed that the newly proposed results encompass both the previously reported unique values and distributions, while previously published BE dispersion does not cover on their own the CO BE distribution reported here.

\subsection{CH$_4$ as adsorbate} 

\begin{figure}[]
    \centering
    \includegraphics[width=0.9\linewidth]{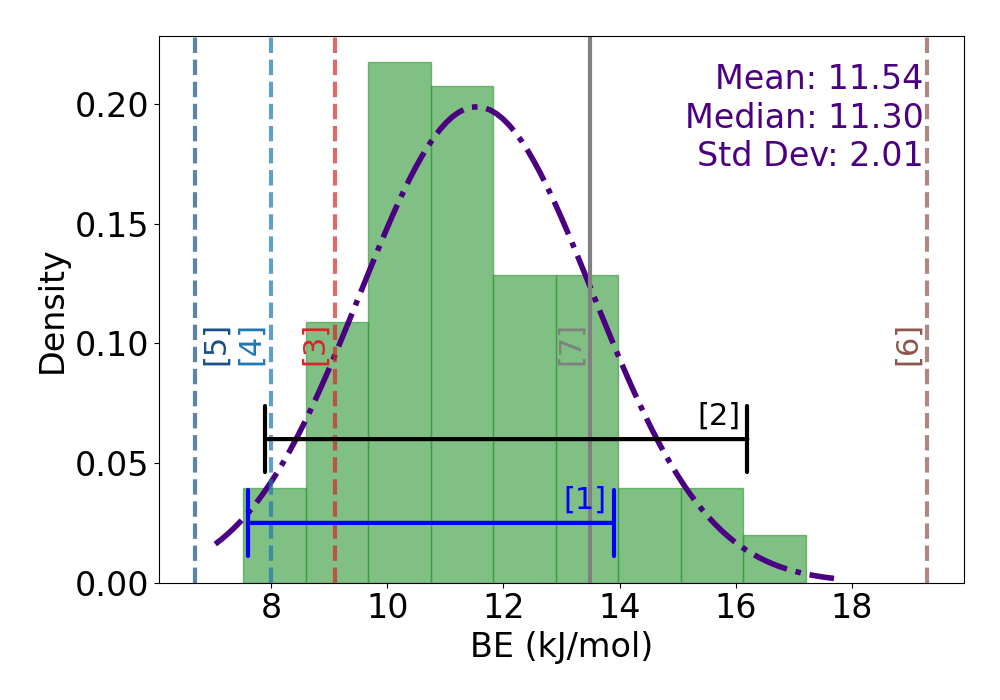} \caption{\label{CH4_distr_corr_BEs_hist}Normalized histogram of sampled binding energies (in kJ/mol) for CH$_4$, fitted with a single component Gaussian profile. Previously reported results of CH$_4$ BE values onto water ice analogs are given on top. See Fig \ref{NH3_distr_corr_BEs_hist} for the meaning of horizontal intervals, vertical dashed lines and the plain gray line. Ref: [1] \cite{Ferrero}; [2] \cite{Smith} \& \cite{He} \& \cite{Penteado}; [3] UMIST; [4] KIDA, [5] \cite{Wakelam}; [6] \cite{Das}; [7] \cite{Ferrero} on crystalline ice.}
    \end{figure}

In the case of CH$_4$, only one starting orientation per cut has been studied. Out of the 100 starting CH$_4$/ice systems, all converged and 6 of them showed imaginary frequencies, among which 4 presenting 2 to max 3 imaginary frequencies in the range $i2$ to $i16$ cm$^{-1}$ have been kept for the statistical analysis. Consequently, 98 CH$_4$ BEs energies have been successfully computed for the sampling of CH$_4$ BE distribution. The normalized BE distribution is shown in Fig. \ref{CH4_distr_corr_BEs_hist}. The mean and median are sensibly close to CO-related values. However, the standard deviation is lower, with a reduced upper limit as compared to CO. This is attributed to the apolarity of CH$_4$. Previously reported single CH$_4$ BE values or dispersion are also reported in Fig. \ref{CH4_distr_corr_BEs_hist}. 

In terms of distributions, the work from \cite{Ferrero}, in which 5 binding sites have been tested, is still considered, as well as the experimental TPD results by \cite{Smith}, \cite{He} and \cite{Penteado}. Both are included in the CH$_4$ BE range reported in this work, and the experimental values cover more than 90\%. However, one should keep in mind that the BE value inferred from TPD experiments greatly depends on the adopted pre-exponential factor for the inversion method. Concerning single values, the KIDA and UMIST references fall in the low-energy part of the distribution from the present work, while the crystalline BE value of \cite{Ferrero} lies in the middle-to-high energy range of our results, as expected. On the other hand, the values of \cite{Wakelam} and \cite{Das} are outside the range of the newly reported distribution. This is probably explained by the size of the water ice model, i.e. a single water molecule in \cite{Wakelam} and a cluster made of up to 6 H$_2$O in \cite{Das}. Such small water systems are not adequately accounting for the converged nature of BEs relative to the size of the ice model, underestimating the ice rigidity, long-range forces and H-bond cooperativity. Together with the convergence analysis presented in Sect. \ref{sec:Discussions}, these results are therefore consistent.

\section{Further discussions}
\label{sec:Discussions}

\subsection{Geometrical and molecular surrounding insights}

\paragraph{(a) NH$_3$ binding behavior.} The inferred NH$_3$ BE distribution displays a bimodal profile. Actually, NH$_3$ is able of forming H-bonds with water, acting as both a donor and an acceptor. More specifically, NH$_3$ is a strong H-bond acceptor, due to the high electron density in the nitrogen lone pair. However, the NH$_3$ H-bond donation capabilities lead to weaker H-bond due to the relatively low polarity of the N-H bond compared to e.g. the highly polar O-H bond in water. As a comparative reference, BEs on H$_2$O-NH$_3$ dimers at CCSD(T)/aug-cc-pVTZ level amount to 27.0 and 9.5 kJ/mol for NH$_3$ acting as acceptor and donor,  respectively, to be compared to 29.7 kJ/mol for H$_2$-H$_2$O H-bonding interaction. With NH$_3$ approaching the surface with different orientations, the binding interactions result in 0 to 3 H-bonds with the underlying water molecules. The case with 4 H-bonds would only be possible if cavity-like binding sites were studied, which is not the case in this paper. 

\begin{figure}[]
    \centering
    \includegraphics[width=0.85\linewidth]{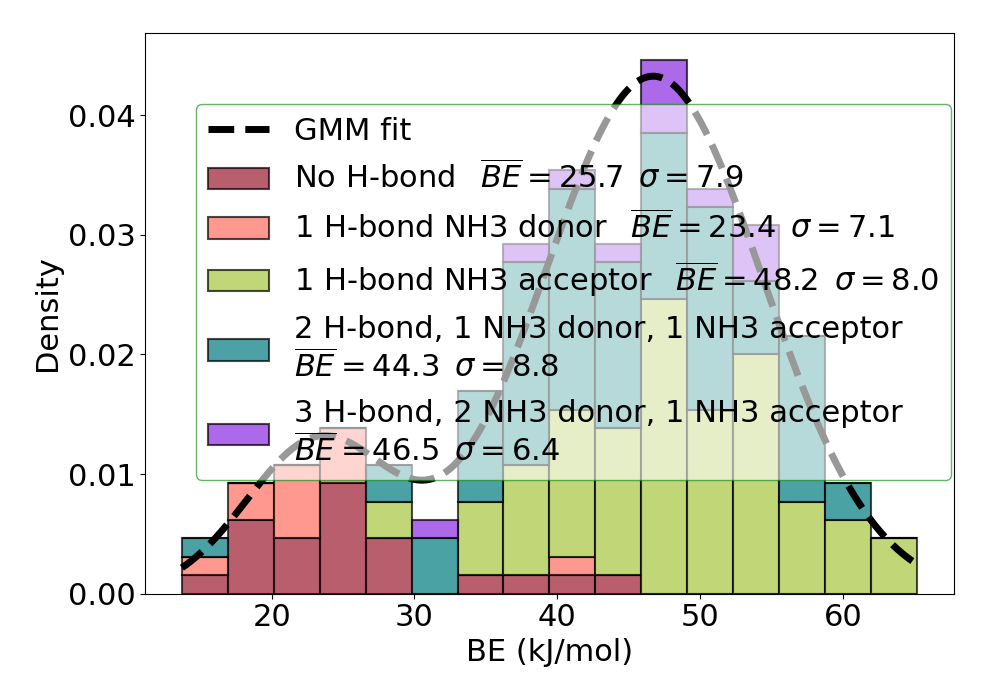}
    \caption{\label{NH3_HB_subdistr_normalized_on_the_gloabl_one}Statistical analysis of subgroup contributions using stacked histograms normalized on the global BE distribution for NH$_3$, each subgroups representing a type of adsorbate-ice H-bond configuration. $\overline{BE}$ and $\sigma$ are the respective subgroup mean BE and standard deviation.}
    \end{figure}

To gain insight into the link between the BE and the number of H-bonds, a statistical analysis of the contributions of subgroups using stacked histograms normalized onto the global NH$_3$ BE distribution has been performed using a custom python script, as presented in Fig. \ref{NH3_HB_subdistr_normalized_on_the_gloabl_one}. Each subgroup is distinguished on the basis of the detailed number of H-bonds, i.e. the number of H-bonds with NH$_3$ as donor (D) or acceptor (A). Note that the H-bond identification, made from a custom python script, is based on purely geometrical cutoffs, with a higher limit onto the distance between heteroatoms fixed at 3.2 $\AA$, and a lower limit for the D-H--A angle defined at 140°. From these results, it appears that the small, left-shifted first peak in the overall distribution is mostly populated by binding configurations implying no H-bond or, at a lower degree, 1 H-bond with NH$_3$ as donor. This is explained by the main implication of non-directional van der Waals interactions, while the other binding configurations imply hydrogen bonding interactions exhibiting a strong directionality as well as a good cooperative property, typically leading to larger BEs. In fact, the dominant contributions to the major peak of the overall distribution come from binding interactions that display at least one H-bond with NH$_3$ as the acceptor. It is also interesting to note that the contributions from cases presenting H-bonding interactions implying only NH$_3$ as donor are very weak. These results are explained by the afore-discussed NH$_3$ H-bonding donation and receiving capabilities. This is further reflected in the mean distance between heteroatoms, being 2.78 \text{\AA} $\pm$ 0.05 \text{\AA} for NH$_3$ acting as acceptor, while it amounts to 3.03 \text{\AA} $\pm$ 0.24 \text{\AA} when it acts as donor. 

The results of this clustering analysis partially align with the machine learning (ML) clustering analysis performed in \cite{Tinacci}, in which the Scikit-learn \citep{scikit-learn} implementation of hierarchical agglomerative clustering has been used, with a predefined number of 2 clusters, using the unreduced BE distribution, the associated electronic BE, the deformation energies, the minimum H-bond distances, and the corresponding angles. This showed that the global distribution can be divided into 2 sub-distributions with distrinct features; weaker adsorption sites are attributed to NH$_3$ acting as H-bond donor, while the stronger binding cases to configurations with NH$_3$ as acceptor. The first peak, highlighted through the ML-clustering approach, is however less pronounced, and is dominated by configurations presenting NH$_3$ as H-bond donor, while the present analysis rather attributes a clear dominance of configurations without any H-bonding interactions. These deviations in the attribution of the characteristic first peak binding features is arguably rationalized by different geometrical cutoffs used for the H-bond identification, as reflected in the typical H-bond angles for H-bond donation configurations (typically lower than 120°, with a min N--H distance between 2.5 and 3.5\text{\AA}) in results from \cite{Tinacci}. 

While hydrogen bonding plays a central role, BE values are likely the result of a intricate interplay of multiple factors, including for instance the local number and proximity of surrounding water molecules, leading to a complex and system-specific energetic behavior.

\begin{figure}[h]
    \centering
    \includegraphics[width=0.85\linewidth]{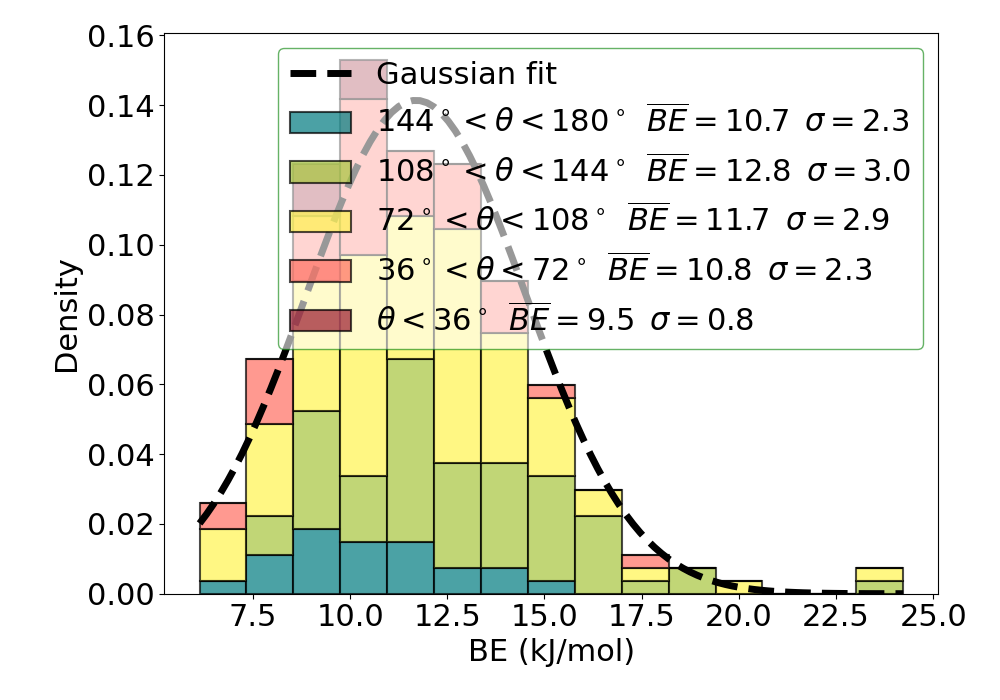} \caption{\label{CO_HB_subdistr_normalized_on_the_gloabl_one}Sub-group contributions of $\theta$ value ranges using stacked histograms normalized on the global BE distribution for CO. $\overline{BE}$ and $\sigma$ account for the respective subgroup mean BE and standard deviation. 
    }
    \end{figure}

\paragraph{(b) CO binding behavior.}
In the case of CO, with the aim of searching for angle-dependent binding behaviors, the orientation of the CO molecule with respect to the reference z-axis has been evaluated by calculating the angle $\theta$ between the molecular axis (C–O vector) and the z reference unit vector. This angle was obtained using the arc-cosine of the normalized dot product between the C–O bond axis and the z-reference vector, yielding values between 0° and 180°. The resulting angle provides insight into the molecular orientation relative to the surface normal, neglecting the substrate roughness; a value of 0° or 180° indicates perfect alignment of the CO axis along the z-direction, respectively, with the carbon or oxygen atom pointing toward the substrate. The intermediate case concerns CO axis perpendicular to the z direction and, therefore, mostly parallel to the substrate surface. As for NH$_3$, a supervised clustering approach has been implemented through a dedicated custom python script, this time discriminating the sub-distributions in terms of $\theta$ values. The analysis of subgroup contributions using stacked histograms normalized onto the global CO BE distribution is presented in Fig. \ref{CO_HB_subdistr_normalized_on_the_gloabl_one}. From these results, it first appears that very few binding configurations involve very low $\theta$ values (0 to 36°), i.e. with the CO axis pointing upward. Furthermore, such CO orientations only cover a narrow low BE range, approximately from 9 to 11 kJ/mol. This suggests that the perpendicular orientation of CO, with the carbon atom facing the surface, results in weaker binding interactions. A slightly higher probability is found for $\theta$ to lie in the 36-72° range, covering a wider range of BE values.
The dominant contributions come from intermediate $\theta$ values (72-108°), although contributions from $\theta$ between 108 and 144° are also significant. Actually, while this subgroup contribution analysis does not lead to strongly differentiated clustering patterns for the different CO orientations, this highlights the proneness of CO to align with the surface. Beside the adsorbate-to-substrate orientation, multiple other factors are expected modulate the computed BEs via intertwined contributions, likely explaining the large coverage of the total CO BE dispersion for those dominant intermediate $\theta$ values. Note also that the highest energy tail of the distribution, i.e. from $\sim$ 15 kJ/mol, is mostly populated by configurations presenting $\theta$ values in the range [108-144°]. This indicates a slight tendency toward stronger interactions for CO configurations where its axis is inclined, pointing the oxygen atoms toward the surface. This is arguably explained by the electronic density and polarity characteristic of the CO bond, with a small dipole moment and partial charges such as the computed CHELPG charges for the isolated CO molecules are +0.017 and -0.017 for the C and O atoms, respectively. In binding configurations that include $\theta$ values in the range [108-144°], the charges are redistributed with $q_C \in [+0.035, +0.255]$; $q_O \in [-0.020, -0.180]$, with oxygen interacting favorably with partially positive regions (hydrogens, with computed CHELPG charges $q_H \in [+0.200 - +0.550]$) from the substrate water molecules. Indeed, ASW surfaces exhibit numerous dangling hydrogens. Very high $\theta$ values (144-180°) have a lower occurrence than intermediate cases, only covering BE ranges from the lower two first thirds of the distribution. This suggests that a situation where oxygen is almost totally pointing downward is less favorable, arguably because of the repulsion effect with the oxygen from the surface. 

\paragraph{(c) CH$_4$ weak-binding case.}
Although CH$_4$ exhibits a relatively high isotropic polarizability (2.12 $\AA^3$) as compared to NH$_3$ (1.72 $\AA^3$) and CO (1.75 $\AA^3$), the absence of permanent dipole moment and the negligible electronegativity difference between carbon and hydrogen leads to non-directional dispersion-dominated weak interactions with the substrate. This therefore precludes the search of any unambiguous configurational trend that meaningfully correlates with the BEs.

\subsection{Analysis of the contributions from $\Delta$ZPE}

\begin{figure}[]
    \centering
    \includegraphics[width=0.9\linewidth]{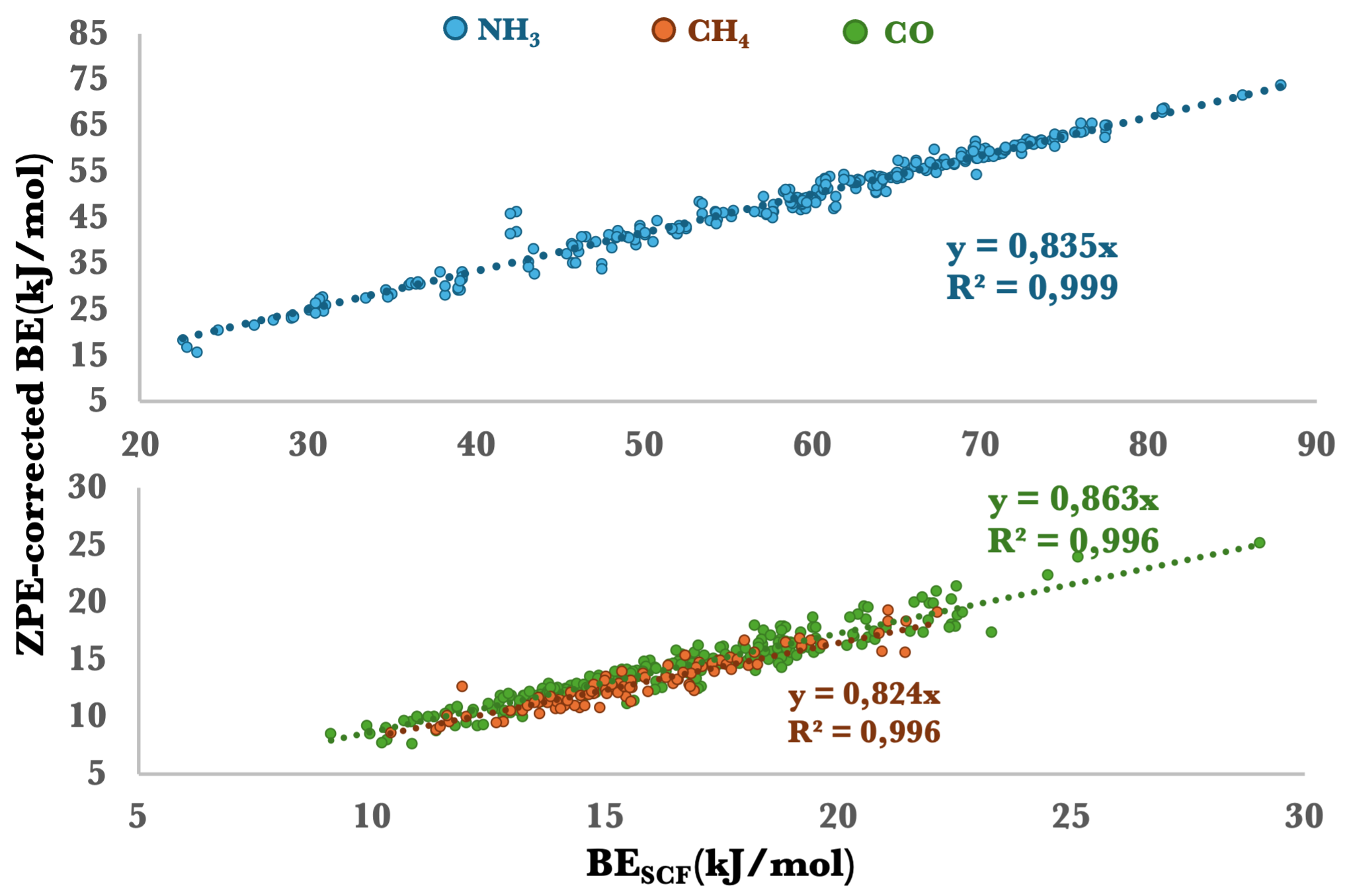} \caption{\label{BEuncorr_vs_BE_ZPE_corr}Linear correlation between the electronic binding energies and the ZPE corrected values (best fit - dashed line) for NH$_3$ (blue), CO (green) and CH$_4$ (orange) on ASW ice}
    \end{figure}

The computation of the ZPE corrections to the binding energies is computationally demanding. Nevertheless, this correction contributes non negligibly to the final inferred BEs. In order to find a robust method to avoid these computations, in line with previous works on the BE distribution of interstellar species on amorphous ices, $BE_{SCF}$ values are compared to the ZPE-corrected BEs. Fig. \ref{BEuncorr_vs_BE_ZPE_corr} provides the corresponding linear correlation for the three adsorbate test cases. Note that the origin has been forced to be equal to 0 to align with previously published results and be able to make rigorous comparisons.

The linear correlation for NH$_3$ is represented by $BE_{ZPE-corrected}$ = 0.835 $BE_{SCF}$ ($R$ = 0.99). In \cite{Germain}, where NH$_3$ BE distribution has been evaluated at full GFN2-xtb level on the same agglomerated cluster of 200 H$_2$O water molecules as in the already largely discussed downstream work by \cite{Tinacci}, a scaling factor of 0.83 ($R$ = 0.98) is found, which is pretty close to the above results. On the other hand, in the periodic framework followed in \cite{Ferrero}, this regression has also been studied, but using results for several adsorbates on crystalline ice. This leads to a a scaling factor of 0.854. This scaling factor has then been applied to the amorphous systems in order to avoid the related resource-demanding ZPE frequency computations on their larger unit cell, representing the amorphous system. 

In the case of CO, the linear correlation between ZPE-corrected BE values and $BE_{SCF}$ demonstrates a very appreciable fit, yielding a scaling factor of 0.865 compared to 0.835 for NH$_3$, thus indicating a 3\% difference. Regarding finally CH$_4$, the scaling factor is reduced to 0.824, to be compared to 0.835 and 0.865 for NH$_3$ and CO, respectively.

The three values are of the same order of magnitude over the three adsorbate test cases, and those three adsorbates are sufficiently different to be representative of the various typical physisorption interactions an adsorbate can involve when binded onto such a surface. It is therefore relevant to consider a universal value for the $\Delta$ZPE scaling factor, applicable to any adsorbate. This approach will align the method in \cite{Ferrero}, though directly applied to the computed frequency data for the amorphous ice model, rather than extrapolated from the crystalline phase. In this context, different methods can be arguably adopted. Firstly, the linear fit between ZPE-corrected BEs and the electronic values including the data for all the three tested adsorbates gives a scaling factor of 0.837. This value is not surprisingly closer to the value from NH$_3$ case since it covers a larger range of BEs. Furthermore, in order to give an equal weighting factor for each adsorbate test case, another approach would be to consider the average of the three values specific to each species, leading to a value of 0.841.

\subsection{Convergence analyses}

The method built throughout this paper is focused toward the unbiased sampling of statistically converged binding energy distributions. A rigorous convergence analysis would, therefore, allow us to objectively comment on the convergence behavior of the datasets. Note that such an analysis has never been applied in the aforementioned previously published works. In that context, a bootstrapping approach has been applied to assess the convergence of the BE distribution statistics with increasing set sizes of binding configurations. For each of the underlying random samplings of the full set, no repetition was allowed. For each set size, 100 bootstrapping sampling runs are performed.

\begin{figure}[]
    \centering
    \includegraphics[width=0.95\linewidth]{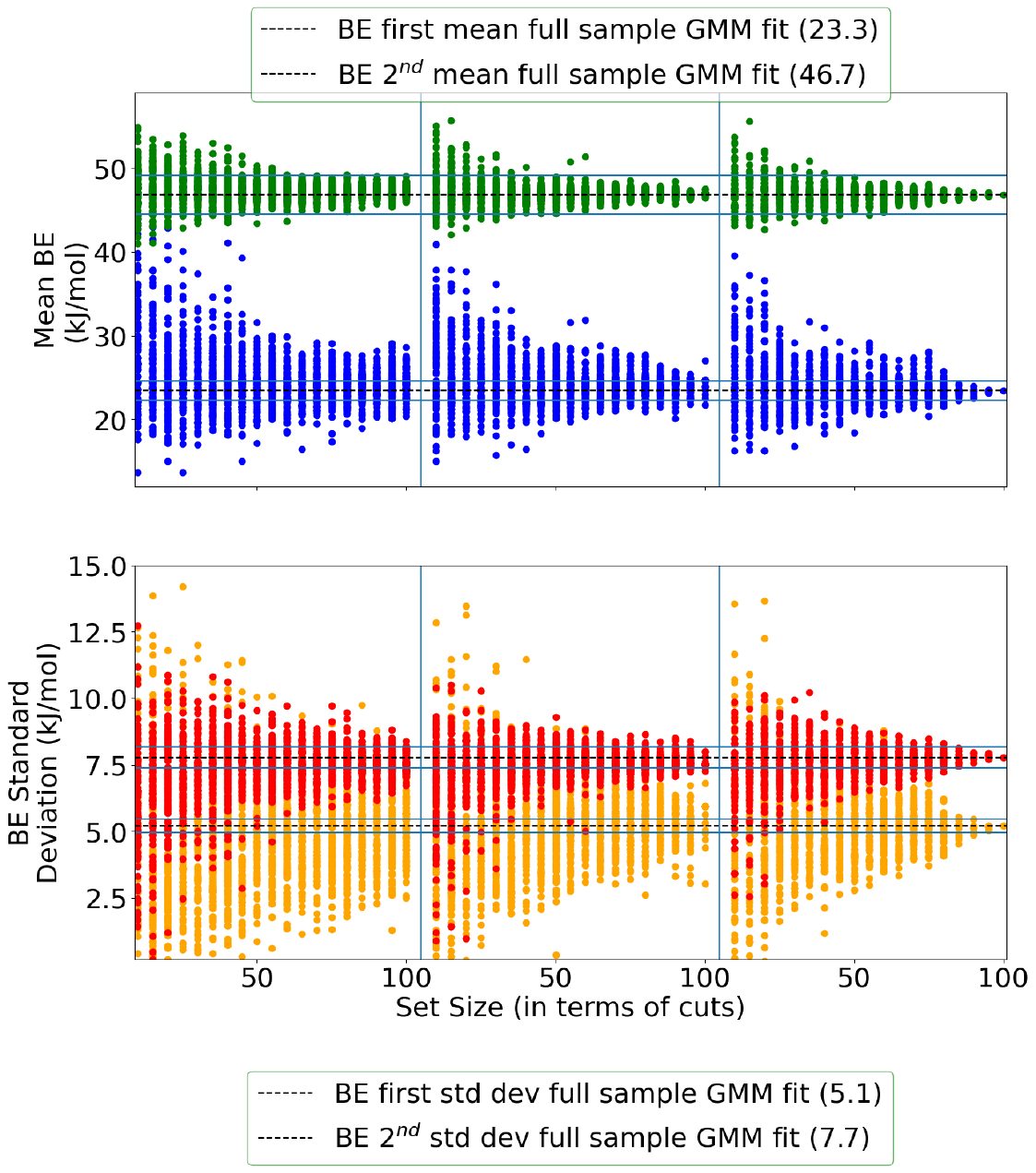} \caption{\label{convergence_analysis_BE_gmm_fit_parameters_stepwise_ads_orientationNH3}Bootstrapping analysis of GMM two components statistics of NH$_3$ BE distribution, with increasing set size. The set size definition depends on the number of picked adsorbate orientation per cut, increasing from 1 to 3 from the left to the right. The dashed line represents the statistics for the full set of data, while the continuous lines encompass a tolerance interval of 5\% around these statistics.}
    \end{figure}

\paragraph{(a) Sampling of the NH$_3$ mixed gaussian BE profile.}
Fig. \ref{convergence_analysis_BE_gmm_fit_parameters_stepwise_ads_orientationNH3} shows the results of such an analysis on NH$_3$ GMM two-component statistics for increasing set sizes, replicating 3 times the bootstrapping method. The set size is sequentially defined for the three runs by increasing the number of selected adsorbate orientation(s) per cut until all of them are considered. More specifically, for the first run, a single BE value corresponding to a given starting adsorbate-to-substrate orientation is randomly chosen out of the 3 studied. For the second run, the BE values from two most diverging starting adsorbate-to-substrate orientations (see Sect. \ref{sec:Methods}) are considered, while all the three orientations per cut are included at last run. At each bootstrapping cycle, for each sampled cut, a binding configuration redundancy analysis is performed when the number of adsorbate orientations considered per cut is greater than or equal to 2. 

From the last bootstrapping run including all the three adsorbate orientations per randomly chosen cut, statistics of both Gaussian components are converged, at least within the convergence criteria, i.e. with a convergence cu-off tolerance interval of 5\% around the mean and standard deviation of the entire distribution. More precisely, convergence is more slowly reached for the low-energy first Gaussian component, intrinsically due to its lower mean and standard deviations, as well as its lower intensity. This is especially true for the standard deviation, with more than 85 cuts required to statistically fall within the defined convergence cutoff. When the number of adsorbate orientations per cut is also considered, increasing the number of starting adsorbate orientations per cut decreases the number of required cuts to reach convergence. Specifically, if only a single orientation per cut is examined, except for the mean for the second and dominant Gaussian component whose convergence is rapid, 100 cuts are insufficient to achieve 5\% accuracy around the standard deviation of this main peak. They are also inadequate for reaching similar precision in both the mean and standard deviation of the first component of the GMM fit. By increasing the number of initial orientations per cut to two, convergence is reached for the second dominant peak, yet remains unmet for the low-energy first peak, especially in terms of standard deviation. Robust convergence of the statistics of the first peak, is only achieved by considering 3 orientations per cut. 

\begin{figure}[]
    \centering
    \includegraphics[width=0.95\linewidth]{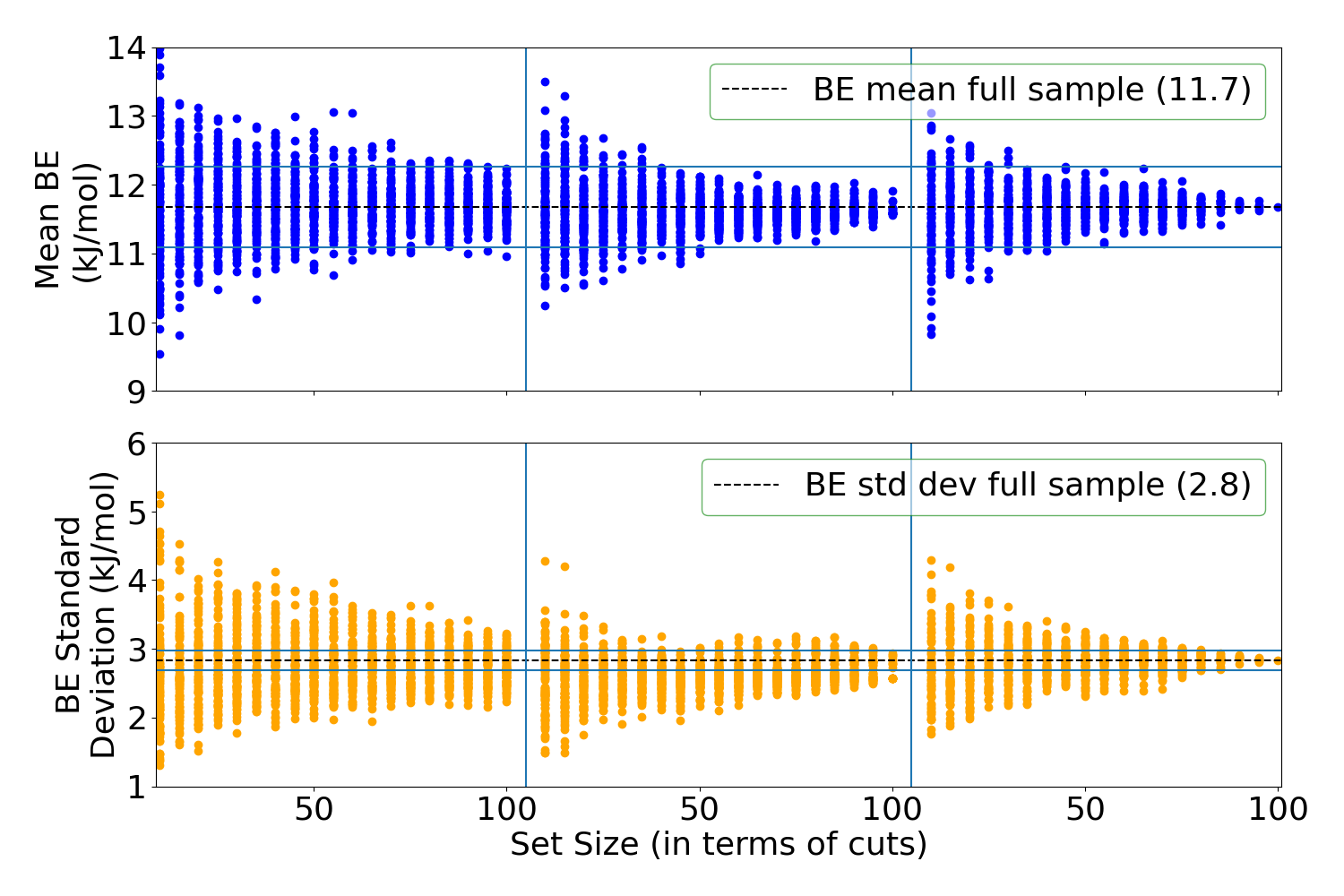} \caption{\label{convergence_analysis_BE_stepwise_ads_orientationCO_considering_in_site_Red}Bootstrapping analysis of mean BE \& standard deviation of CO on LDA ice, with increasing set size. The set size definition depends on the number of picked adsorbate orientation per cut, increasing from 1 to 3 from the left to the  right.}
    \end{figure}

\paragraph{(b) CO BE dispersion.}
The bootstrapping analysis discussed above has been applied to the global mean and standard deviation for sets with increasing set sizes, considering in-site binding configuration redundancy reduction. The results are given in Fig. \ref{convergence_analysis_BE_stepwise_ads_orientationCO_considering_in_site_Red}. Starting with the rightmost bootstrapping run, including all three geometries per sampled cut, the results indicate converged global statistics, with the mean naturally converging faster than its standard deviation due to the very small magnitude of the latter. To achieve convergence for both within the defined cut-off, a set size of at least 80 cuts is required. However, the convergence criterion for the standard deviation could be slightly relaxed, depending on the desired level of precision. On the other hand, similar to NH$_3$, increasing the number of considered adsorbate orientations per cut leads to faster convergence in terms of cuts, resulting in greater standard deviation stability. Although the convergence (within the defined criterion) of the standard deviation is not reached when only one adsorbate orientation is randomly chosen per cut, it is almost reached after 100 cuts when the two most differing starting adsorbate geometries are considered. However, as just stated, the very low value of the standard deviation intrinsically results in slower convergence. To optimize the accuracy-computational cost balance based on the convergence of global statistics, studying the two most diverging orientations of CO per cut could consequently be sufficient, once again depending on the level of precision one expects to reach.
    
\begin{figure}[h]
    \centering
    \includegraphics[width=0.92\linewidth]{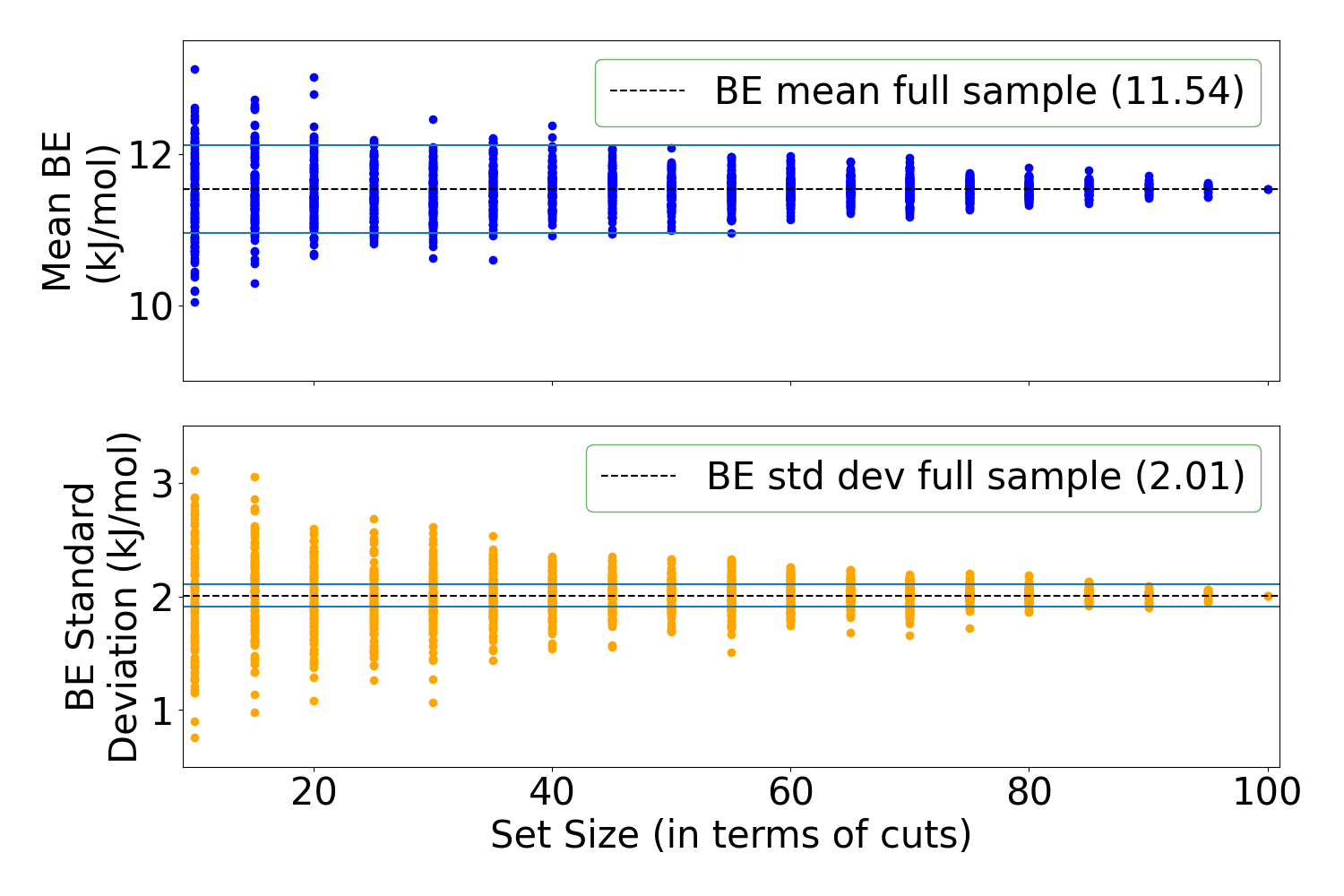} \caption{\label{convergence_analysis_BE_CH4}Bootstrapping analysis of mean BE \& standard deviation of CH$_4$ on LDA ice, with increasing set size. The set size is defined by the number of picked random hemispherical systems. }
    \end{figure}

\paragraph{(c) CH$_4$ BE dispersion.}
Results are given in Fig.\,\ref{convergence_analysis_BE_CH4}. Despite the lower number of sampled binding configurations due to the single starting orientation of CH$_results4$, the convergence appears to be well attained under the consideration of at least 85 cuts. The sampling of a larger surface area is thereby not required. 

\subsection{Astrochemical Implications} \label{sec:astro_implic}

The three adsorbate test cases were chosen for their distinctive typical interactions with water surfaces, as well as their significance in interstellar ices. In addition to water, CO, NH$_3$ and CH$_4$ are among the main components of such interstellar water-rich ices \citep{Hama_Watanabe_review_surf_processes,Oberg_layered_model}. Thoroughly studying their BE distribution is, therefore, necessary for an accurate kinetic modeling. In the case of NH$_3$, the double-Gaussian profile is anticipated to have a non-negligible impact on the solid-phase chemistry of cold molecular clouds, where weak binding configurations may enable surface diffusion/desorption, whereas stronger binding cases inhibit it. Actually, as discussed in \cite{Sipila,Tinacci}, the gaseous abundance of ammonia in cold prestellar clouds \citep{Crapsi}, e.g. the L1544 prestellar core, cannot be attributed to the gas-phase chemistry. In fact, NH$_3$ is expected to be completely frozen out in icy matrix in typical prestellar core temperatures, that is, 7 K in the deep interior of L1544 \citep{Crapsi,Keto}. However, \cite{Sipila} found that a chemical desorption of less than 1\% of the NH$_3$ formed on dust grains is sufficient to reproduce the gas-phase abundances measured in L1544. In that context, they explored the possibility that the ammonia BE is lower than the typically considered high value (45.7 kJ/mol), and therefore tested BE values of 24.9 kJ/mol, as suggested in \cite{Kamp}, and references therein, as well as an ad hoc very low value of 8.3 kJ/mol. The latter led to an overestimation of the ammonia gas phase abundance by orders of magnitude. However, as commented in \cite{Tinacci}, considering the inferred BE distribution rather than a single BE value may improve the agreement with the observed gas-phase ammonia abundance. Quantitatively, the first peak of the NH$_3$ BE distribution reported in this work, spanning BE range between $\sim$ 15 and 30 kJ/mol, covers 16.3\% of the total distribution. This suggests a chemical desorption efficiency even lower than the upper limit of 1\% proposed in \cite{Sipila}. However, as discussed in \cite{Sipila}, the gas-phase abundance of NH$_3$ in such a prestellar core may also be arguably explained by the late freezing out of CO molecules in the deep interior of the cloud. This alters the ice composition and arguably decreases the NH$_3$ depletion degree due to weaker binding interactions with the CO-dominated ice. This is still a matter of debate, with the layering of the ice featuring an uppermost CO apolar ice in the late evolution of molecular clouds \citep{Oberg} being contested by the models in \citep{Kouchi,Kouchi1}. If the layering process is effective, we suggest that a mix of both the first BE peak in the NH$_3$ BE distribution on ASW with weak chemical desorption efficiency, coupled with the change of the ice composition at late cloud evolutionary stages, is likely explaining the gas-phase abundance of ammonia in a prestellar core such as L1544. Further modeling is needed to verify this suggestion.

\section{Conclusions}
\label{sec:Conclusions}

In this paper, we presented a revised methodology for the BE distribution inference of interstellar species present in water-rich amorphous icy mantle in dense clouds and young protostellar cores, using NH$_3$, CO and CH$_4$ as adsorbate test cases. This methodology relies on the following key points:

    \begin{itemize}
        \item[i.] The structural reliability of the MD-modeled ASW ice analogue, validated through comparisons to previously reported structural data;

        \item[ii.] The design of the applied ONIOM-2 scheme for the BE computation aimed to be applicable to a wide range of interstellar species. It involves several confirmation steps, with

            (a) a benchmark on the DFT functional for the description of the model zone. Using CCSD(T) as reference, B3LYP-D3(BJ)/6-311+G(d,p) stands out as the best accuracy-to-computational needs balance; 
            
            (b) a convergence analysis of the the real system size effect on the computed BE values, using GFN2-xtb as low-level method and a model zone size of 8 $\AA$. This showed that the influence of the real system size depends on the adsorbate as well as the binding site. Over the three adsorbate test cases, a real system size of 16 $\AA$ provides the best compromise. This analysis also demonstrated that a too low real system size may lead to artifacts in the inferred BE values; 
            
            (c) an analysis of the GFN2-xtb performances (with respect to a DFT-based treatment) under growing real system sizes, showing that GNF2-xtb is suitable for this study, at least within the chosen real system size of 16 $\AA$;

            (d) a final retro-verification of the size of the ONIOM layers, confirming the choice of a model zone of 8 $\AA$ within a real system of 16 $\AA$.  

        \item[iii.] A twofold binding configuration sampling, exploring the diversity both in terms of locally distinct binding sites, and of the local roughness of the PES. The first source of diversity is sampled through the use of a regularly-spaced sampling grid above the surface, defining the position of the adsorbate barycenter. On the other hand, the local roughness of the PES is studied by using multiple starting adsorbate-to-substrate orientations for each binding site. The number of starting orientations is defined by the adsorbate symmetry. This sampling method is designed to ensure the statistical convergence of the inferred BE distributions, as proven through bootstrapping convergence analysis onto the respective mean and standard deviation of the BE distributions.
    \end{itemize}

Following this modeling framework, the BE distribution of NH$_3$, CO and CH$_4$ have been successfully inferred, showing converged statistics. More specifically, results showed that the NH$_3$ BE distribution is best represented by a double Gaussian profile, with a first low-energy peak dominated by configurations without any H-bonding interaction, or to a lower extent with 1 H-bond with NH$_3$ as donor. This first peak may be of great astrochemical significance, as further commented on the L1544 protostellar core, in line with the discussions in \cite{Tinacci}. Moreover, the convergence analyses also demonstrated that, in the cases of NH$_3$ and CO for which the symmetry justified the study of multiple adsorbate-to-substrate orientations, considering only one random adsorbate orientation per cut is not sufficient to reach convergence with the full distribution statistics. This is the case evenly when two adsorbate orientations are chosen per cut, at least in the case of NH$_3$ and its two GMM components. In the case of CO, depending on the defined convergence cut-offs, considering 2 adsorbate orientations per cut can be considered as sufficient. In terms of comparison with previously reported results, results from this work encompass all previously reported values, except in the case of CH$_4$ for which two unique BE values, inferred from very small systems, fall outside the dispersion range reported here. This is probably explained by a too limited size of the ice model used to compute these two values.

As a perspective, the method presented in this paper is expected to be applied to other adsorbates relevant to the interstellar medium. The primary focus should be on CO$_2$, H$_2$CO and CH$_3$OH, i.e. other important constituents of molecular clouds and young protostellar ices \citep{Hama_Watanabe_review_surf_processes}. Furthermore, the impact of the ASW model density onto the resulting BE distributions will be the subject of a subsequent paper.

\begin{acknowledgements}
     We thank the anonymous referee for his/her time, and for his/her fair and positive report. M. Groyne thank the Belgian National Fund For Scientific Research (F.R.S.-FNRS) for the Research Fellow fellowship. These computations were performed on the computers of the ‘‘Consortium des Équipements de Calcul Intensif (CÉCI)’’ (https://www.ceci-hpc.be), including those of the ‘‘UNamur Technological Platform of High-Performance Computing (PTCI)’’ (https://www.ptci.unamur.be) and those of the Tier-1 supercomputer of the Fédération Wallonie-Bruxelles, for which we gratefully acknowledge the financial support from the FRS-FNRS, the Walloon Region, and the University of Namur (Conventions No. U.G006.15, U.G018.19, U.G011.22, RW1610468, RW/GEQ2016, RW1117545, and RW2110213).  
\end{acknowledgements}

\bibliographystyle{aa} 
\bibliography{references}

\begin{appendix}

\section{Ice model - Further validation of the structural reliability}
\label{Appendix_deep_comparison_structural_ice_features}

\begin{figure}[h]
    \centering
    \includegraphics[width=1\linewidth]{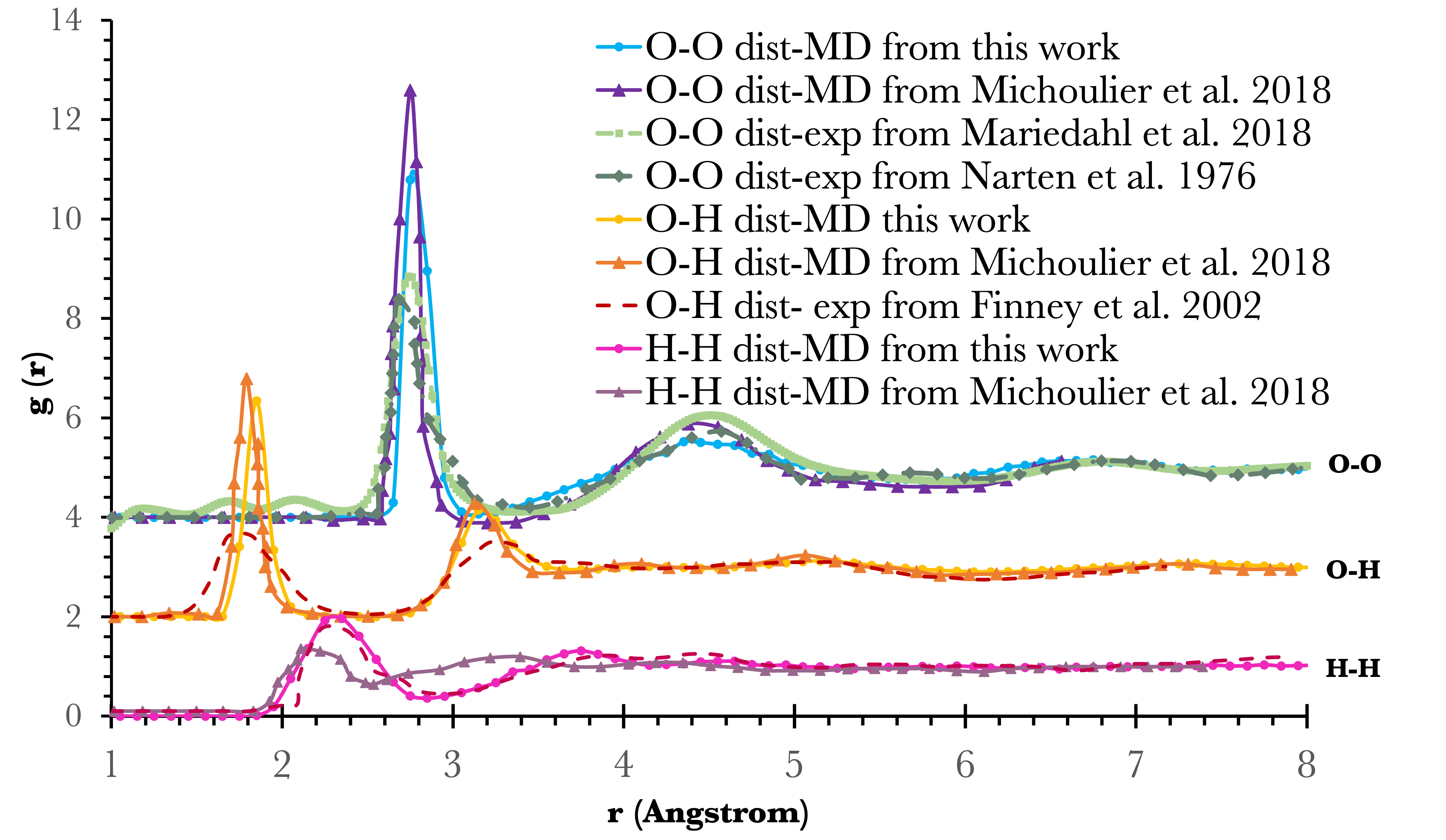}
    \caption{Intermolecular O-O, O-H and H-H RDFs from this work, \cite{Mariedahl} and \citet{Michoulier} (with \citealt{Narten_HDA_density} and \citealt{Finney} data used in \citealt{Michoulier}).}
    \label{g(r)_large_comparison}
    \end{figure}

A broader comparison has been made to other MD simulations \citep{Michoulier} and experimental results \citep{Narten_HDA_density,Finney,Mariedahl} to address all pairs' distribution functions. The results are given in Fig. \ref{g(r)_large_comparison}. From a general point of view, the simulated ices of \citet{Michoulier} and this work agree very well. Moreover, the modeled data from \cite{Michoulier} display the same small qualitative differences discussed above with respect to the experimental curves as our simulations. This match is fairly appreciable despite the different MD simulation protocols, where \cite{Michoulier} modeled their ASW ice analogues from conversion of crystalline water ice. This provides a robust warranty to the aforementioned comment on the expected negligible impact of the ice building protocol on the overall structure of the ice model and subsequent inferred BEs. 

We also checked the number of H-bonds per water molecules. This can be studied from (i) the $g_{O-H}(r)$, in combination to its integrated form $n_{O-H}(r)$, or (ii) through analysis of the mean number of H-bonds per water molecule/oxygen atom, as provided by the VMD (alpha 1.9.4 version) visualization and analysis software \cite[]{VMD}. $n_{O-H}(r)$ and $n_{O-O}(r)$ display a clear plateau at 3.96-4 in the distance range corresponding respectively to the inter-molecular $g_{O-H}(r)$ and $g_{O-O}(r)$ first peak. This further highlights the almost perfect tetrahedrality of the H-bond network. Concerning the mean number of H-bonds per water molecule, it amounts to 3.97, tending towards the net value of 4 in crystalline water ice. 

These results are fully consistent with the discussion in the main text and in line with theoretical considerations. This further validates the suitability of the use of the TIP4P/2005 Force Field in our scheme.

\section{Benchmark onto the DFT functional for the model system description - detailed results}
\label{Appendix_HL_functionnal_details}

\begin{figure*}[]
    \centering
    \includegraphics[width=0.88\linewidth]{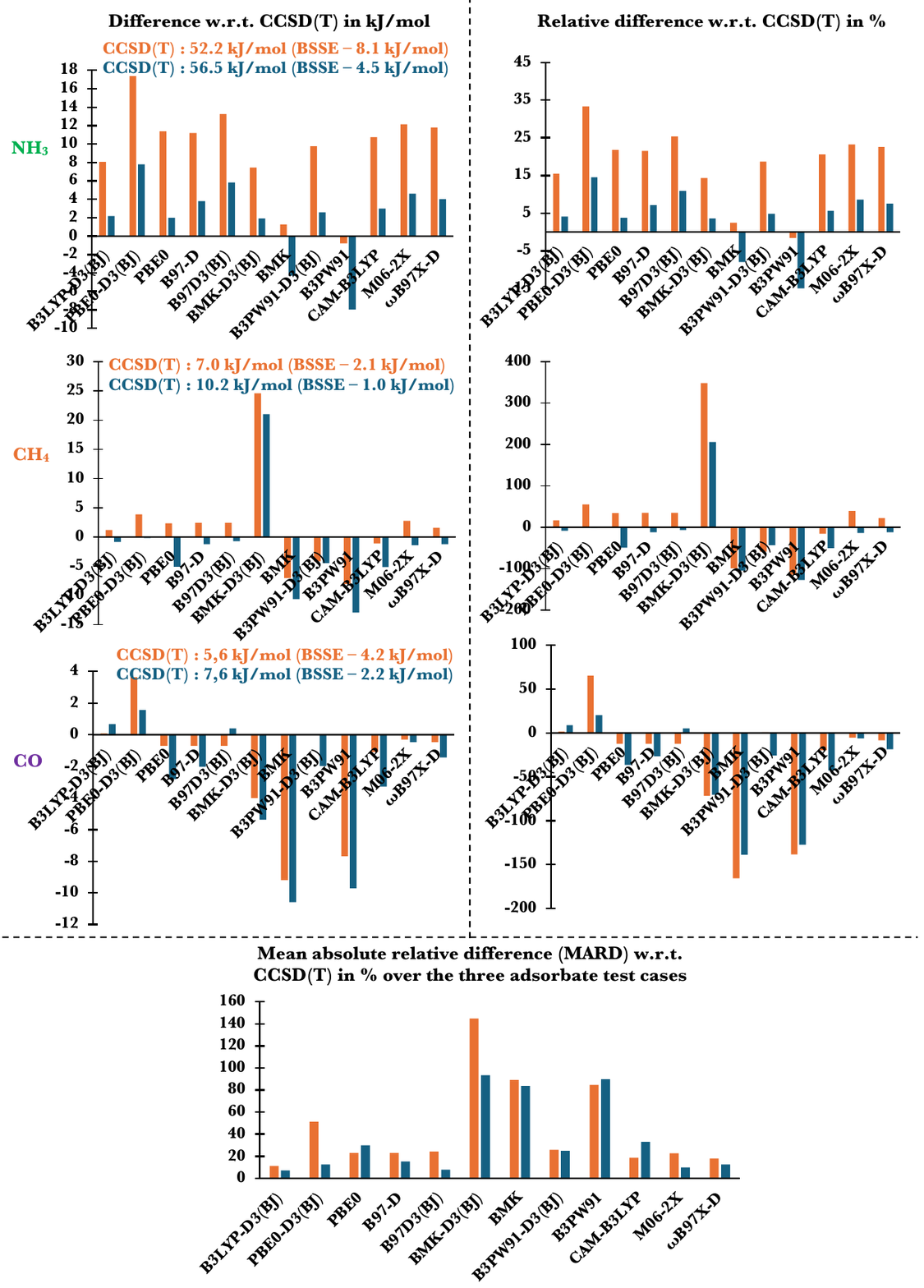}
    \caption{\label{Graphe-ONIOM_Scheme_design_high_level_functionnal_benchmark} DFT-functional benchmark on the tetrameric structures; orange and blue series account respectively for results with 6-311+G(d,p) and aug-cc-pVTZ as basis set - left column:  difference with respect to CCSD(T) value. CCSD(T) reference binding energy and BSSE correction values are given explicitly; right column: relative difference with respect to CCSD(T) value in \%. Graphs are going by pairs (raw), with the first raw being attributed to NH$_3$, the second to CO and the third to CH$_4$. The last raw gives the mean absolute relative differences over the three adsorbate.}
    \end{figure*}

Fig. \ref{Graphe-ONIOM_Scheme_design_high_level_functionnal_benchmark} provides detailed results for the DFT functional benchmark for the model zone description. Results obtained with 6-311+G(d,p) are represented in orange, while blue results stand for QM schemes with the extended correlation-consistent aug-cc-pVTZ as basis set. The first three raws gives results for each adsorbate, while the last raw compares the mean absolute relative differences (MARD) with respect to CCSD(T) over the three adsorbate test cases. As we can see, the tendencies obtained with 6-311+G(d,p) as basis set are qualitatively retrieved with aug-cc-pVTZ, harboring reduced discrepancies with respect to CCSD(T) values. This is principally due to the magnitude of the BSSE correction onto the reference value, lowered in the case of the aug-cc-pVTZ basis set and its additional diffuse and polarization functions. These diffuse functions are associated to small exponents extending over a much larger spatial domain. Their spatial extension leads to weak overlapping with many other basis functions, introducing very small eigenvalue to the overlap matrix slowing down the SCF convergence. Considering our ultimate goal to extensively sample an amorphous surface, with relatively large ice substrate (see Sect. \ref{Low_Lev_Design_Sect}),  B3LYP-D3(BJ)/6-311+G(d,p) level of theory offers the best compromise for a proper accuracy/computational cost balance for the Model zone treatment. Note that the choice of 6-311+G(d,p) as high-level basis set is further validated in Appendix \ref{Appendix_high_level_BS_against_growing_LowLevel}. 

\newpage
\section{Stability of the benchmark on the real system size under extended high-level basis set}
\label{Appendix_high_level_BS_against_growing_LowLevel}

\begin{figure}[h]
    \centering
    \includegraphics[width=1\linewidth]{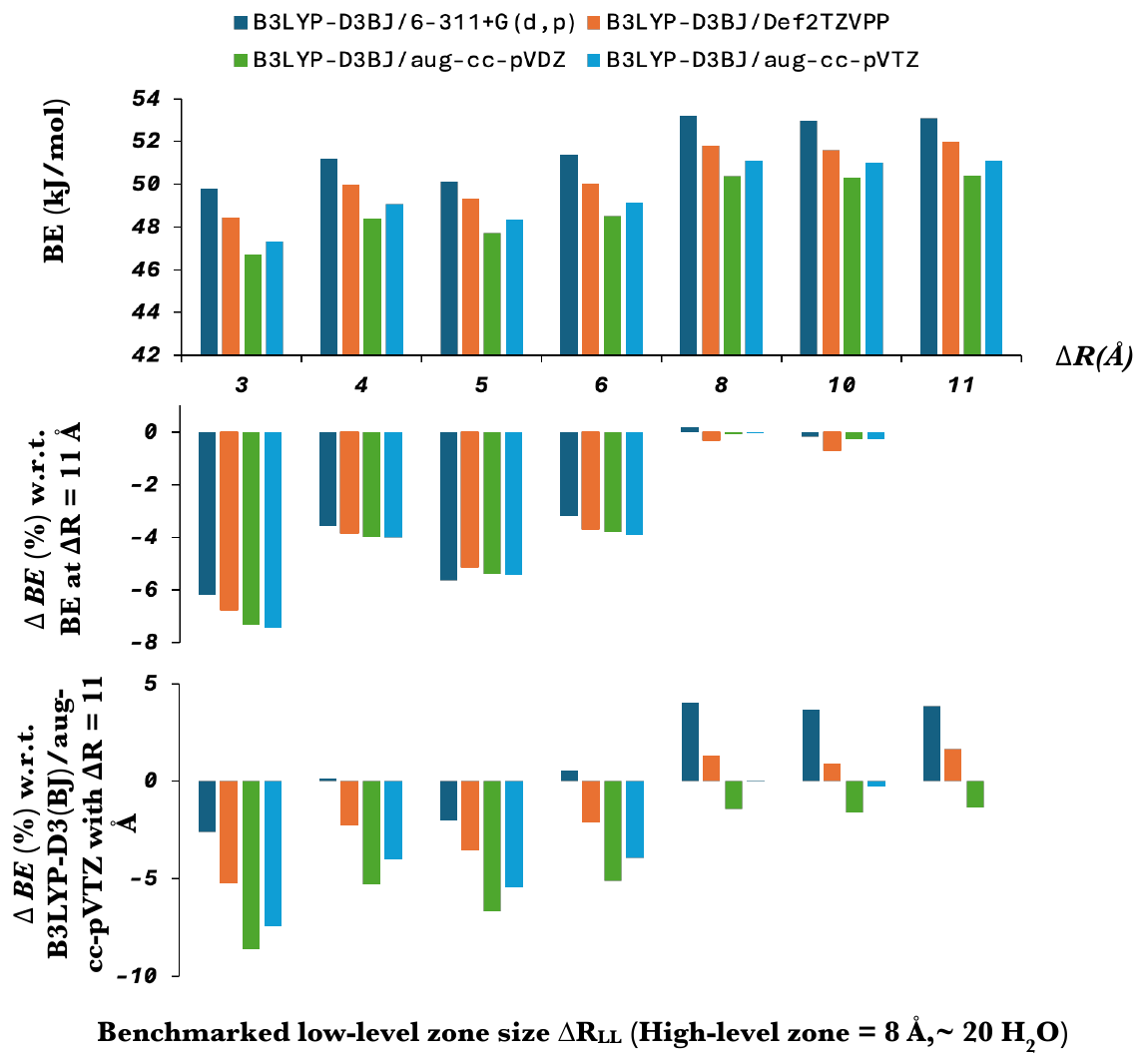}
    \caption{\label{NH3_Graphe-ONIOM_Scheme_design_HL_BasisSet_analysis_and_LL_bench} (top panel) NH$_3$ binding energies on a first tested binding site for a growing $\Delta R_{Low Level}$, going from 3 to 11 \AA, for a total radius between 11 and 19 \AA. 4 different basis set for the high-level treatment are tested, as given in the legend of the plot, within an ONIOM(B3LYP-D3(BJ)/basis set:xtb) framework; (Middle panel) Relative BE difference with respect to BE at $\Delta R_{Low Level}$ equal to 11 \AA, each series being treated independently; and (bottom panel) Relative BE discrepancies relative to BE at B3LYP-D3(BJ)/aug-cc-pVTZ high level of theory, with $\Delta R_{Low Level}$ equal to 11 \AA. }
    \end{figure}

\begin{figure}[]
    \centering
    \includegraphics[width=1\linewidth]{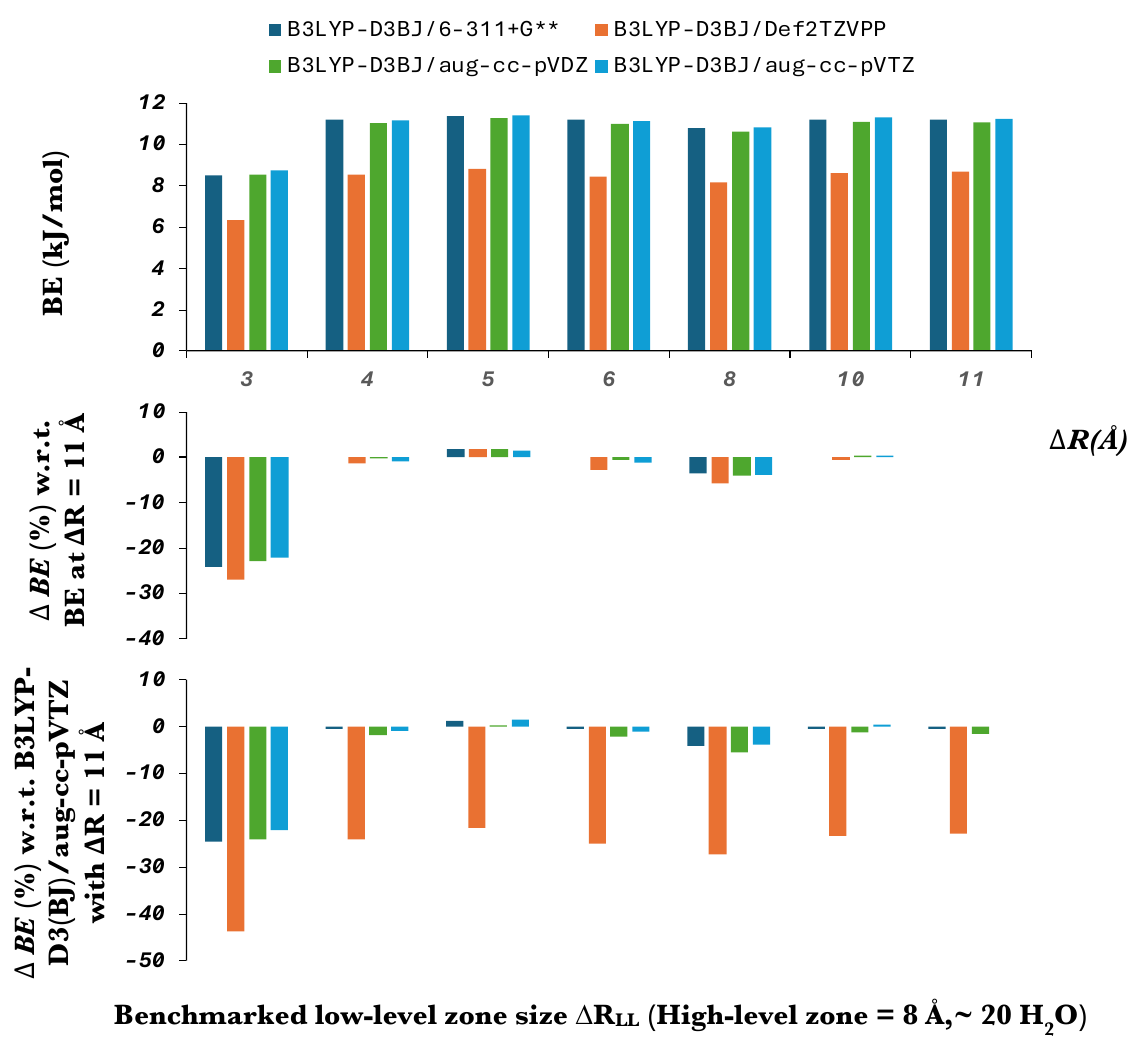}
    \caption{\label{CO_Graphe-ONIOM_Scheme_design_HL_BasisSet_analysis_and_LL_bench} CO binding energies on a first tested binding site for a growing $\Delta R_{Low Level}$, going from 3 to 11 \AA, for a total radius between 11 and 19 \AA. 4 different basis set for the high-level treatment are tested, as given in the legend of the plot, within an ONIOM(B3LYP-D3(BJ)/basis set:xtb) framework. See legend Fig. \ref{NH3_Graphe-ONIOM_Scheme_design_HL_BasisSet_analysis_and_LL_bench} for the description of each panel.  }
    \end{figure}

\begin{figure}[]
    \centering
    \includegraphics[width=1\linewidth]{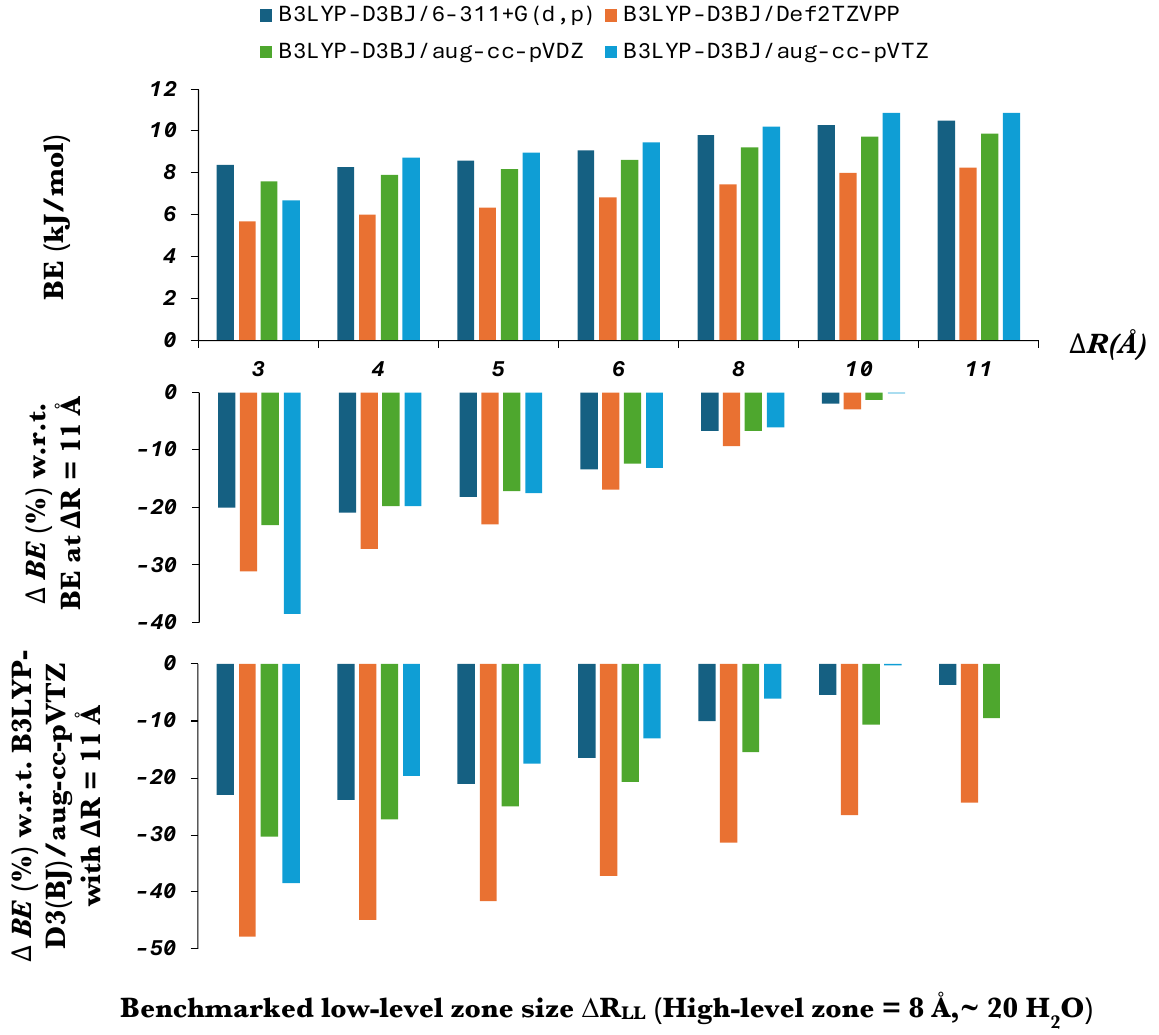}
    \caption{\label{CH4_Graphe-ONIOM_Scheme_design_HL_BasisSet_analysis_and_LL_bench} CH$_4$ binding energies on a first tested binding site for a growing $\Delta R_{Low Level}$, going from 3 to 11 \AA, for a total radius between 11 and 19 \AA. 4 different basis set for the high-level treatment are tested, as given in the legend of the plot, within an ONIOM(B3LYP-D3(BJ)/basis set:xtb) framework. See legend Fig. \ref{NH3_Graphe-ONIOM_Scheme_design_HL_BasisSet_analysis_and_LL_bench} for the description of each panel. }
    \end{figure}

\begin{figure}[]
    \centering
    \includegraphics[width=1\linewidth]{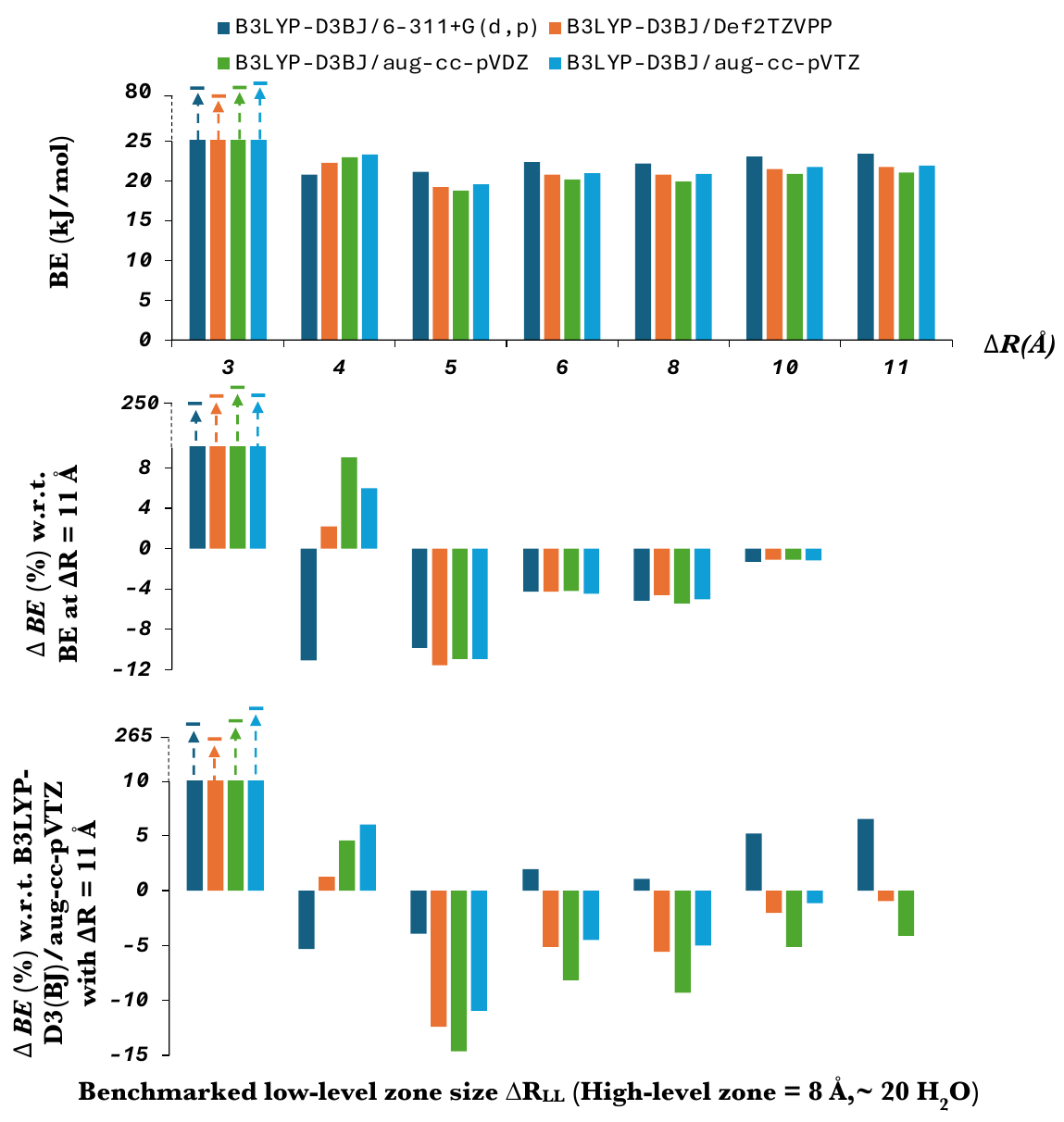}
    \caption{\label{NH3_2nd_Graphe-ONIOM_Scheme_design_HL_BasisSet_analysis_and_LL_bench} NH$_3$ binding energies on a second tested binding site for a growing $\Delta R_{Low Level}$, going from 3 to 11 \AA, for a total radius between 11 and 19 \AA. 4 different basis set for the high-level treatment are tested, as given in the legend of the plot, within an ONIOM(B3LYP-D3(BJ)/basis set:xtb) framework. See legend Fig. \ref{NH3_Graphe-ONIOM_Scheme_design_HL_BasisSet_analysis_and_LL_bench} for the description of each panel. }
    \end{figure}

In order to check the stability of the benchmark results on the real system size under extended high level basis sets, results obtained within an ONIOM(B3LYP-D3(BJ)/6-311+G(d,p):xtb) framework are compared to results obtained with 3 other basis sets for the model zone description, i.e. def2-TZVPP, aug-cc-pVDZ and aug-cc-pVTZ. Results are gathered in Fig. 
\ref{NH3_Graphe-ONIOM_Scheme_design_HL_BasisSet_analysis_and_LL_bench}, \ref{CO_Graphe-ONIOM_Scheme_design_HL_BasisSet_analysis_and_LL_bench}, \ref{CH4_Graphe-ONIOM_Scheme_design_HL_BasisSet_analysis_and_LL_bench} and \ref{NH3_2nd_Graphe-ONIOM_Scheme_design_HL_BasisSet_analysis_and_LL_bench} respectively for NH$_3$, CO and CH$_4$ on the first binding site, and NH$_3$ on the second binding site. 

In the case of NH$_3$, focusing on the middle panel of Fig. \ref{NH3_Graphe-ONIOM_Scheme_design_HL_BasisSet_analysis_and_LL_bench}, the convergence between BEs and the real system size arises, for each series, from a $\Delta R_{Low Level}$ of 8 $\AA$. In the case of Fig. \ref{NH3_2nd_Graphe-ONIOM_Scheme_design_HL_BasisSet_analysis_and_LL_bench}, every BE series also display consistent convergence trend with respect to the real system size, rather appearing at 10 $\AA$ but already appreciable from $\Delta R_{Low Level}$ of 6-8 $\AA$. In terms of $\Delta BE$ w.r.t. results obtained with a model system treatment at B3LYP-D3(BJ)/aug-cc-pVTZ high level of theory at the largest real system size, these are relatively weak and very stable at system size-BE convergence for the first NH$_3$ binding site, as seen in the bottom panel from Fig. \ref{NH3_Graphe-ONIOM_Scheme_design_HL_BasisSet_analysis_and_LL_bench}. In the case of the second site, the bottom panel of Fig. \ref{NH3_2nd_Graphe-ONIOM_Scheme_design_HL_BasisSet_analysis_and_LL_bench} shows that, at $\Delta R_{Low Level}$ in the 6-8 $\AA$ range, the ONIOM(B3LYP-D3(BJ)/aug-cc-pVTZ:xtb) result at largest system size is best represented by the ONIOM(B3LYP-D3(BJ)/6-311+G(d,p):xtb). Increasing $\Delta R_{Low Level}$ further, the difference between ONIOM(B3LYP-D3(BJ)/aug-cc-pVTZ:xtb) and ONIOM(B3LYP-D3(BJ)/6-311+G(d,p):xtb) increases to $\sim$ 5\%. These results for the two tested binding sites of NH$_3$ therefore reinforce our choice for the real system size, with $\Delta R_{Low Level}$ equal to 8 $\AA$. 

In the case of CO and CH$_4$, the convergence between the real system size and the BE values is also consistent among the different series. In terms of $\Delta BE$ relative to model system treatment at B3LYP-D3(BJ)/aug-cc-pVTZ high level of theory at the largest real system size, def2-TZVPP provides results with larger deviations as compared to results with 6-311+G(d,p), aug-cc-pVDZ and aug-cc-pVTZ as high-level basis set. This is likely explained by the absence of diffuse functions in def2-TZVPP, required in the context of the description of the model zone, especially for the weak interaction involved in CO and CH$_4$ binding behaviors. More specifically, at system size-BE convergence, the converged BE value with aug-cc-pVTZ as high-level basis set is best matched by results with 6-311+G(d,p) as high-level basis set, with relative discrepancies lower than 10\% for CH$_4$ and 5\% for CO. 

From an overall point of view, one therefore gets consistent conclusions in terms of convergence between the real system size and the inferred BEs for the different high-level ONIOM frameworks over the three adsorbate test cases. At convergence, we observed an overall good match of the results obtained with ONIOM(B3LYP-D3(BJ)/6-311+G(d,p):xtb) and ONIOM(B3LYP-D3(BJ)/aug-cc-pVTZ:xtb), with systematic additional diffuse and polarization functions for the correlation-consistent aug-cc-pVTZ basis set, recognized as particularly relevant in weakly interacting systems. This indicates that 6-311+G(d,p) is suitable to capture the essential electronic structure features relevant to this study and constitutes the final foundation for the applied high level of theory, i.e. B3LYP-D3(BJ)/6-111+G(d,p).

\section{GFN2-xtb performance for the low-level description - stability of the performance against growing real system size}
\label{Appendix_xtb_performances_check}

While the convergence of the BEs against the real system size has been studied, it is nevertheless worth to question the reliability of the used low-level method, namely the semi-empirical quantum mechanical GFN2-xtb algorithm from the Grimme group \cite[]{GFN2-xtb,xtb}, as well as the stability of this reliability itself under growing real system sizes. For that purpose, BEs (corrected for $\Delta$ZPE and BSSE) from ONIOM(B3LYP-D3/6-311+G(d,p) : GFN2-xtb) are compared to results from ONIOM(B3LYP-D3/6-311+G(d,p): B3LYP-D3/ growing basis set) for each of the three adsorbate test cases on the firstly studied binding site, with growing hemisphere radius. The tested low-level basis sets correspond to 6-31G, 6-31G(d), 6-31G(d,p), 6-31+G and 6-31+G(d). 

    \begin{figure}[]
    \centering
    \includegraphics[width=1\linewidth]{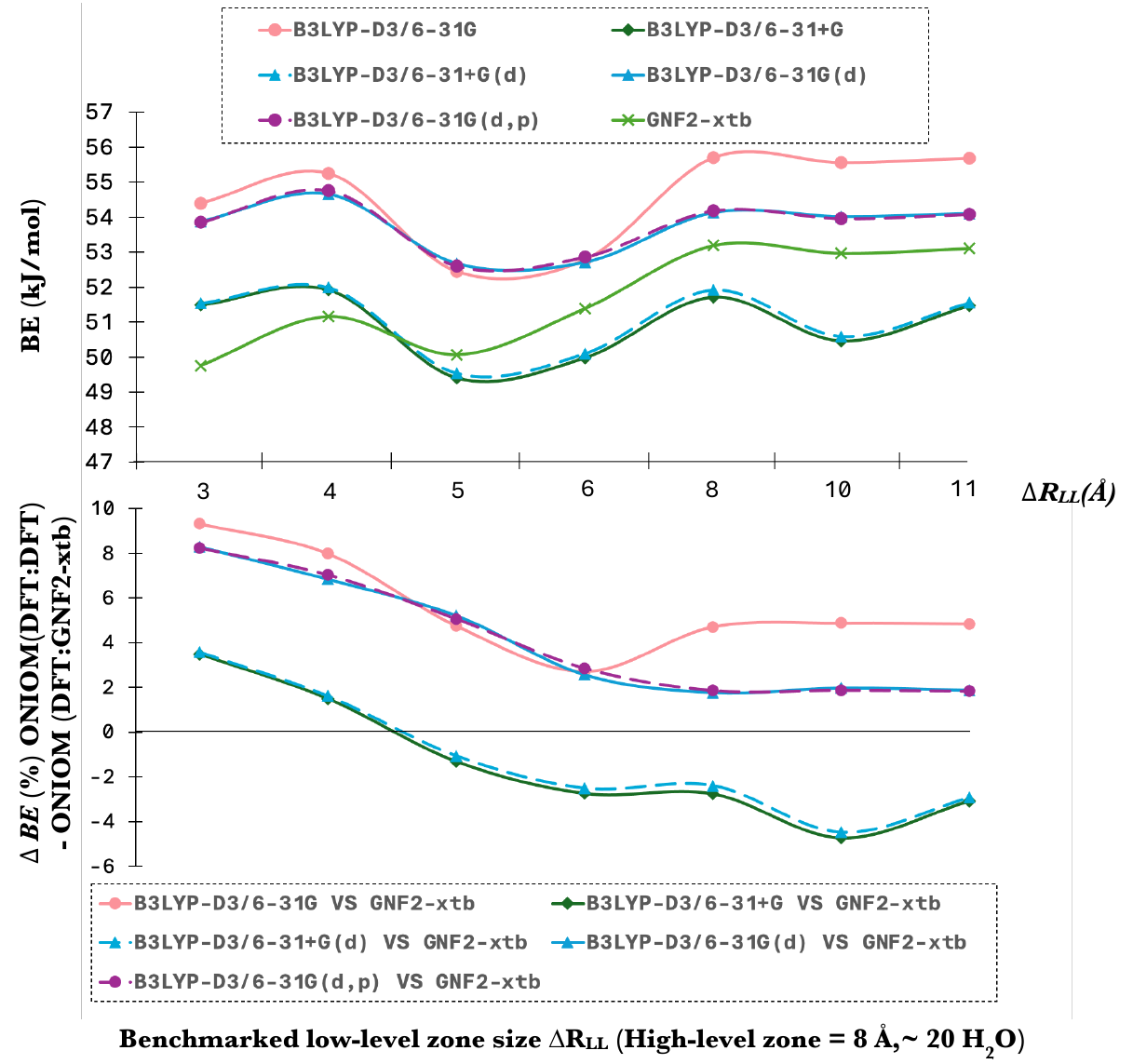}
    \caption{\label{NH3_Graphe-ONIOM_Scheme_design_results-xtb_perf_test}Top panel - NH$_3$ BE for a growing $\Delta R_{LL}$ and various low-levels of theory. The different series hold for a computational scheme following ONIOM(B3LYP-D3/6-311+G(d,p): x), where x, the low-level method, accounts for either for B3LYP-D3/growing basis set (see legend), or GFN2-xtb; Bottom panel - residue graph representing the relative difference between ONIOM(B3LYP-D3/6-311+G(d,p) : GFN2-xtb) and ONIOM(B3LYP-D3/6-311+G(d,p): B3LYP-D3/growing basis set). Note that legends only explicit the low-level method of each series, for the sake of clarity. Lines do not represent any interpolated physical trends, they only serve as guides to highlight the trend and facilitate the reader view of the plots.}
    \end{figure}

\begin{figure}[]
    \centering
    \includegraphics[width=1\linewidth]{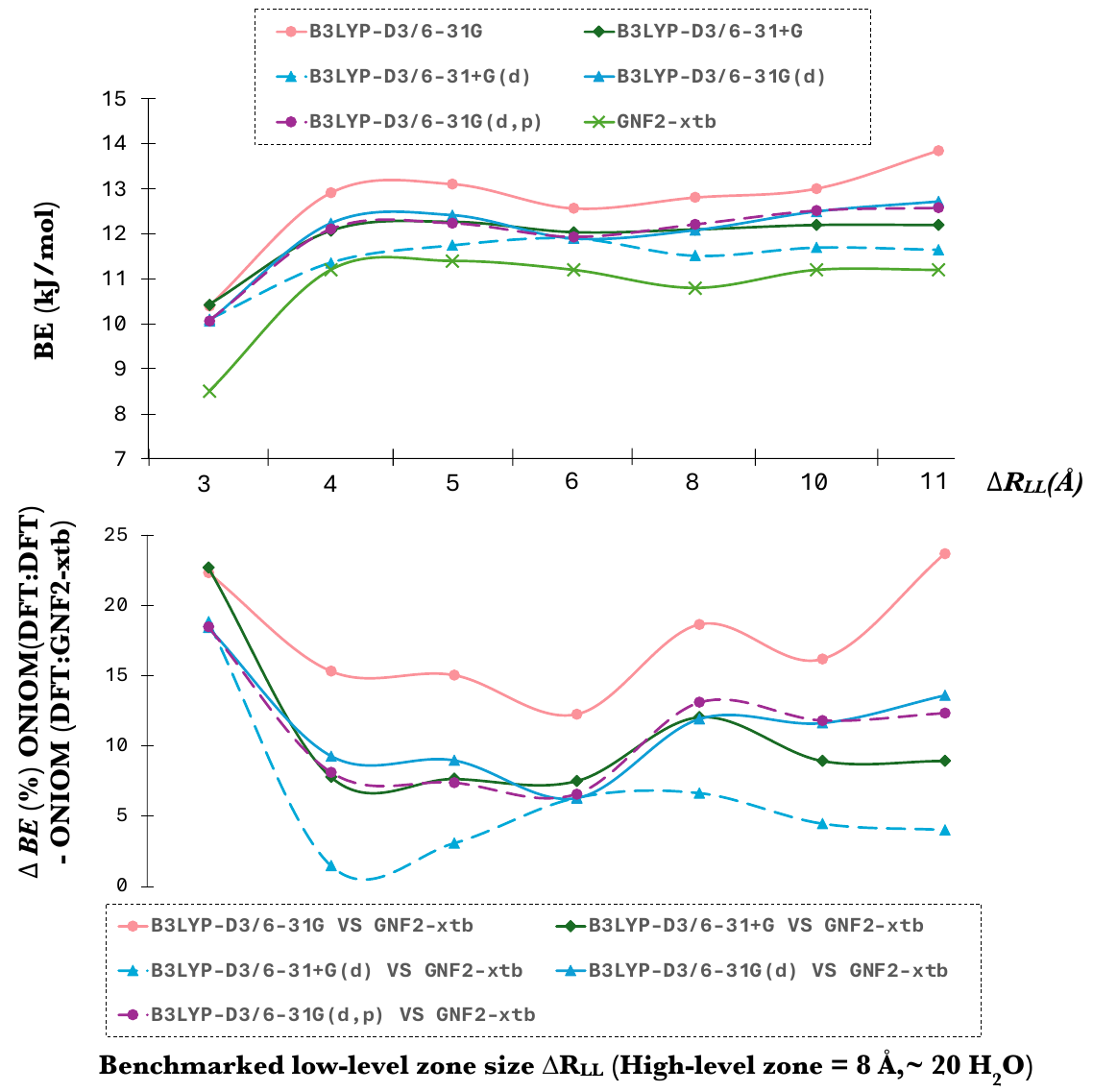}
    \caption{\label{CO_Graphe-ONIOM_Scheme_design_results-xtb_perf_test}CO binding energies for a growing $\Delta R_{LL}$ and various low-levels of theory. See legend of Figure \ref{NH3_Graphe-ONIOM_Scheme_design_results-xtb_perf_test} for further details.}
    \end{figure}

In the case of NH$_3$ (Fig. \ref{NH3_Graphe-ONIOM_Scheme_design_results-xtb_perf_test}), BE values globally converge in each series from $\Delta R_{LL}$ of 8 \AA, as does the ONIOM(B3LYP-D3/6-311+G(d,p) : GFN2-xtb) series alone. The convergence of the series with a low-level treatment at B3LYP-D3(BJ)/6-31+G and B3LYP-D3(BJ)/6-31+G(d) is less pronounced but still appreciable, with a maximum deviation of 1.5 kJ/mol for values of $\Delta R_{LL}$ between 8 and 11 \AA. Note that the observed BE oscillations in this $\Delta R_{LL}$ range may arguably be due to geometrical artifact at the hemisphere boundaries, but no reason has been found to explain the absence of this effect in the other series, or for CO-related series (see Fig. \ref{CO_Graphe-ONIOM_Scheme_design_results-xtb_perf_test}). Concerning the $\Delta$BE between ONIOM(B3LYP-D3/6-311+G(d,p) : GFN2-xtb) and ONIOM(B3LYP-D3/6-311+G(d,p): B3LYP-D3/ growing basis set), at system-size convergence, the ONIOM(B3LYP-D3/6-311+G(d,p) : GFN2-xtb) results best represent the ONIOM(B3LYP-D3/6-311+G(d,p) : B3LYP-D3/6-31G(d)) and ONIOM(B3LYP-D3/6-311+G(d,p) : B3LYP-D3/6-31G(d,p)) frameworks. Still, the relative differences with B3LYP-D3(BJ)/6-31+G and B3LYP-D3(BJ)/6-31+G(d) remain below 5\% at BE-system size convergence. It indicates that xtb is efficient in capturing essential features of low-level processing, at least in this context of NH$_3$ binding dominated by H-bonding interactions.

    \begin{figure}[]
    \centering
    \includegraphics[width=1\linewidth]{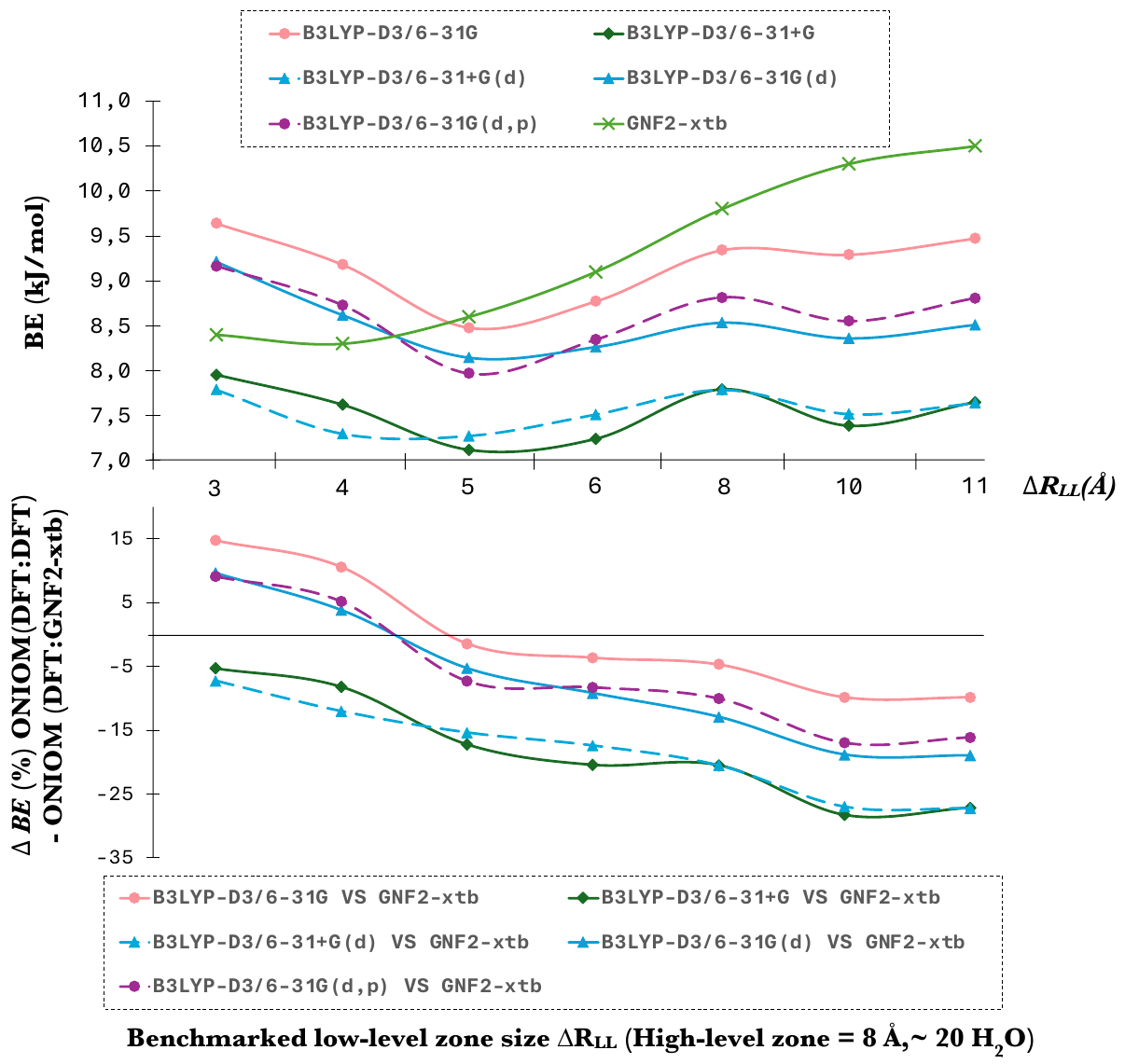}
    \caption{\label{CH4_Graphe-ONIOM_Scheme_design_results-xtb_perf_test}CH$_4$ binding energies for a growing $\Delta R_{LL}$ and various low-levels of theory. See legend of Figure \ref{NH3_Graphe-ONIOM_Scheme_design_results-xtb_perf_test} for further details.}
    \end{figure}

Regarding CO-related series in Fig. \ref{CO_Graphe-ONIOM_Scheme_design_results-xtb_perf_test}, as for ONIOM(B3LYP-D3/6-311+G(d,p) : GFN2-xtb), all series quickly converge with the low-level size, at $\Delta R_{LL}$ of 4 \AA, except for ONIOM(B3LYP-D3/6-311+G(d,p) : B3LYP-D3/6-31G). In terms of relative discrepancies w.r.t. a semi-empirical xtb low-level treatment, these are globally below 15\% at $\Delta R_{LL}$ larger than 4 \AA, except for ONIOM(B3LYP-D3/6-311+G(d,p) : B3LYP-D3/6-31G). This indicates that the cheaper semi-empirical GFN2-xtb method is well parameterized to describe such a dipole-dipole binding interaction. 

Concerning finally the trends obtained for CH$_4$ (Fig. \ref{CH4_Graphe-ONIOM_Scheme_design_results-xtb_perf_test}), xtb performances are weaker, but further validate the choice of a $\Delta R_{LL}$ of 8 $\AA$ in order to make a good trade-off between the stability of the BE values with respect to the system size, as well as the reliability of the low-level method and its consistency with increasing low-level sizes. More specifically, the convergence of the BE series against growing low-level size appears at a $\Delta R_{LL}$ larger than 6 to 8 $\AA$ for full ONIOM(DFT:DFT) frameworks, before the previously discussed convergence of the ONIOM(DFT:xtb) results. The trend followed by the latter series actually differs from those with systems treated at the ONIOM(DFT:DFT) level, while relative trends between ONIOM(DFT:DFT) series themselves are globally consistent. $\Delta$BE between ONIOM(DFT:DFT) and ONIOM(DFT:xtb) frameworks increases at convergence of the ONIOM(DFT:DFT) series, from a $\Delta R_{LL}$ larger than 6 to 8 $\AA$. This was not observed for the CO and NH$_3$ cases. This arguably points out a cooperative effect from the environment not well reproduced by GFN2-xtb. This can likely be attributed to the very weak interactions involved in the CH$_4$ binding interactions, dominated by dispersion terms. In the case of NH$_3$ and CO, stronger interactions were involved, imposing more constraints on the structure of the system. Their cooperative and extended long-range effects are better captured by the semi-empirical xtb method. Quantitatively speaking, the relative differences with respect to a GFN2-xtb low-level treatment, at convergence of the BE series with growing $\Delta R_{LL}$ in full DFT ONIOM schemes, amount to maximum $\sim$ 20\% at $\Delta R_{LL}$ of 8 \AA, while it increases to $\sim$ 30\% at worst at larger real system sizes compared to low-level treatment including diffuse functions. 

Note also that the reliability of GFN2-xtb has already been widely validated for general purposes, as in the original \cite{GFN2-xtb} paper. In systems similar to our study, it has also been evaluated in \cite{Germain} and \cite{Tinacci}. Our results are therefore complementary to their analysis, providing additional method-specific insights into the stability of xtb performance against growing $\Delta R_{LL}$.

\section{Automated script for an unbiased binding configurations sampling}
\label{Appendix_automated_sampling_script}

An automated python script has been written for the generation of all inputs needed for the unbiased sampling of the binding configurations of the three adsorbate test cases onto the modeled ASW surface, and the monitoring of each computations required to obtain the final BEs (geometry optimizations, frequency computations and BSSE evaluation). More specifically, the water box at the end of the last MD quenching phase is replicated in the x-y plane, taking advantage of the PBC. A grid of 10 by 10 equally spaced hemispheric centers is placed above the surface. The grid spacing is 4 \text{\AA}, and avoids redundancy between adjacent adsorption sites. This has been validated through RMSD analysis as further discussed later one. These 100 points constitute the central points for the definition of each of the studied hemispheres based on distance discrimination ($R_{hem}$ = 16 \text{\AA}), taking advantage of the PBC in the x and y directions. The first starting guess of the adsorbate barycenter is then placed at 2 Å in z from each central point. This allows the sampling of spatially distinct binding sites, and therefore accounts for the complexity of the surface typology. The adsorbate-to-substrate orientation is then defined for each grid point. The number of starting orientations depends on the adsorbate symmetry; for NH$_3$, 3 starting orientations have been studied, with (i) with the three hydrogens towards the surface, (ii) with the 3 hydrogen flipped in the opposite direction in z, and (iii) with the plane passing by the three hydrogens being perpendicular to the surface. In the case of CO, 3 orientations have also been sampled, with (i) CO axis parallel to the surface, (ii) CO axis parallel to the normal to the surface, with O towards the surface, and (ii) CO axis parallel to the normal to the surface, with C pointing towards the surface. This defines the most representative/differing orientations, and seems to be the best compromise to sample as much as possible the different possible binding configurations while keeping a reasonable number of systems to be optimized and studied. On the other hand, regarding the symmetry of CH$_4$, only one orientation has been studied for this adsorbate. This enables the sampling of the roughness of the local PES within each locally distinct binding site. The z-position of the adsorbate barycenter is then refined such that at least one adsorbate atom is at a distance lower than 3 Å from one water molecule from the substrate. Atoms are finally grouped by ONIOM-2 layers based on distance discrimination, avoiding any cut bonds at the boundaries, and input for  geometry optimization at ONIOM(B3LYP-D3/6-311+G(d,p) : GNF2-xtb) of theory are written. Tight convergence criteria\footnote{These are defined with the following cutoffs: Maximum force $< 1.5 \times 10^{-4}\ \mathrm{Hartree/bohr}$, RMS force $< 1.0 \times 10^{-4}\ \mathrm{Hartree/bohr}$, maximum displacement $< 6.0 \times 10^{-4}\ \mathrm{bohr}$, and RMS displacement $< 4.0 \times 10^{-4}\ \mathrm{bohr}$.
} are applied for the model zone, while default SCF parameters\footnote{$10^{-6}\ \mathrm{Hartree}$ for the energy and $10^{-4} \mathrm{e}$ for the wavefunction.} are used for GFN2-xtb low-level treatment. Each input is saved for each adsorbate, and subsequently submitted. The monitoring of each job is then ensured automatically through crontab scripts aimed at checking the status of each job (in run, finished normally due to reached convergence, or finished before convergence), and re-launch the appropriate computations until optimization, vibrational frequency (for zero-point energy correction) and BSSE correction computations are completed.  

\section{RMSD analysis of binding configurations (including nearby water molecules)}
\label{Appendix_RMSD_analysis}

As discussed in the main text, binding site redundancy reduction has been applied before addressing the inferred BE distributions. More specifically, following our BE inference scheme, there are two potential source of redundancy (in term of BE and RMSD of the surrounding nearest water molecules) in the binding behavior of a given adsorbate, 

    \begin{itemize}
        \item Redundancy between binding configurations on juxtaposed binding sites/hemispheric cuts. All in all, the inclusion or reduction of this first source of redundancy simply depends on the refinement of the sampling grid. For instance, in the method proposed by \citet{Tinacci}, reduction was required. In our method, this has been avoided by a proper choice of our grid spacing for the binding site starting position sampling, i.e. 4 \text(\AA). We however still checked if no redundancy effectively stands out from the RMSD analysis of juxtaposed binding configurations, as justified in the following paragraphs.
        
        \item Redundancy between binding configuration for different starting orientation of the adsorbate on a given hemispheric cut. This concerns the case of NH$_3$ or CO; if different starting orientations converge to the same local minimum, this has been reduced to the unique consideration of the distinct binding configuration per site. In the followings, this redundancy reduction will be called in-site binding configuration redundancy reduction.  
    \end{itemize}

In-site binding configuration redundancy analyses are performed under the following criteria: two binding configurations are considered to be redundant and need to be reduced if $\Delta$BE$_{ij}$ < 0.05.<BE>$_{ij}$ kJ/mol\footnote{This scaling of the cutoff on the $\Delta$BE with the mean BE of the compared configurations is more consistent than a fixed $\Delta$BE cutoff, e.g. 1 kJ/mol. It actually ensures the same strictness of the criterion whatever the adsorbate and characteristic BEs.} and RMSD < 1 $\AA$ as our redundancy cutoff, where <BE>$_{ij}$ is the mean BE for i and j orientations. 

From these principles, starting from the working hypothesis that the sampled surface (40x40 \text(\AA)) combined to our grid density for the starting guess of adsorption site are statistically representative of a realistic interstellar cold environment ice, and knowing the total number of adsorption sites for a given typical grain surrounded by an icy mantle, the frequency count of each adsorption sites can be directly retrieved.

For each system, we considered as configuration to be analyzed for the RMSD computation\footnote{Note that all RMSD analyses presented in this Appendix have been performed using the open source RMSD computation project available at \url{https://github.com/charnley/rmsd}.} the adsorbate plus the $x$ nearest\footnote{Nearest H$_2$O are defined by ordering the minimum distance between each water molecule and adsorbate atoms. In other words, for each water molecule from the substrate, the distance between each of its constitutive atoms and the atoms from the adsorbate is computed, and the minimum distance is kept for the ordering process of nearby water molecules.} water molecules from the adsorbate, $x$ being equal to the mean number of nearby water molecules within an interaction zone defined by a radius of 3.2 \text{\AA} from each adsorbate atom. Moreover, rotation has been allowed for procedure of minimization of the RMSD. This allows to have a constant number of coordinates to be compared through RMSD analysis, while avoiding as much as possible geometrical artifacts and properly quantify the surface structural diversity in terms of binding site. 

\subsection{NH$_3$ related RMSD analyses}

\begin{figure}[]
    \centering
    \includegraphics[width=1\linewidth]{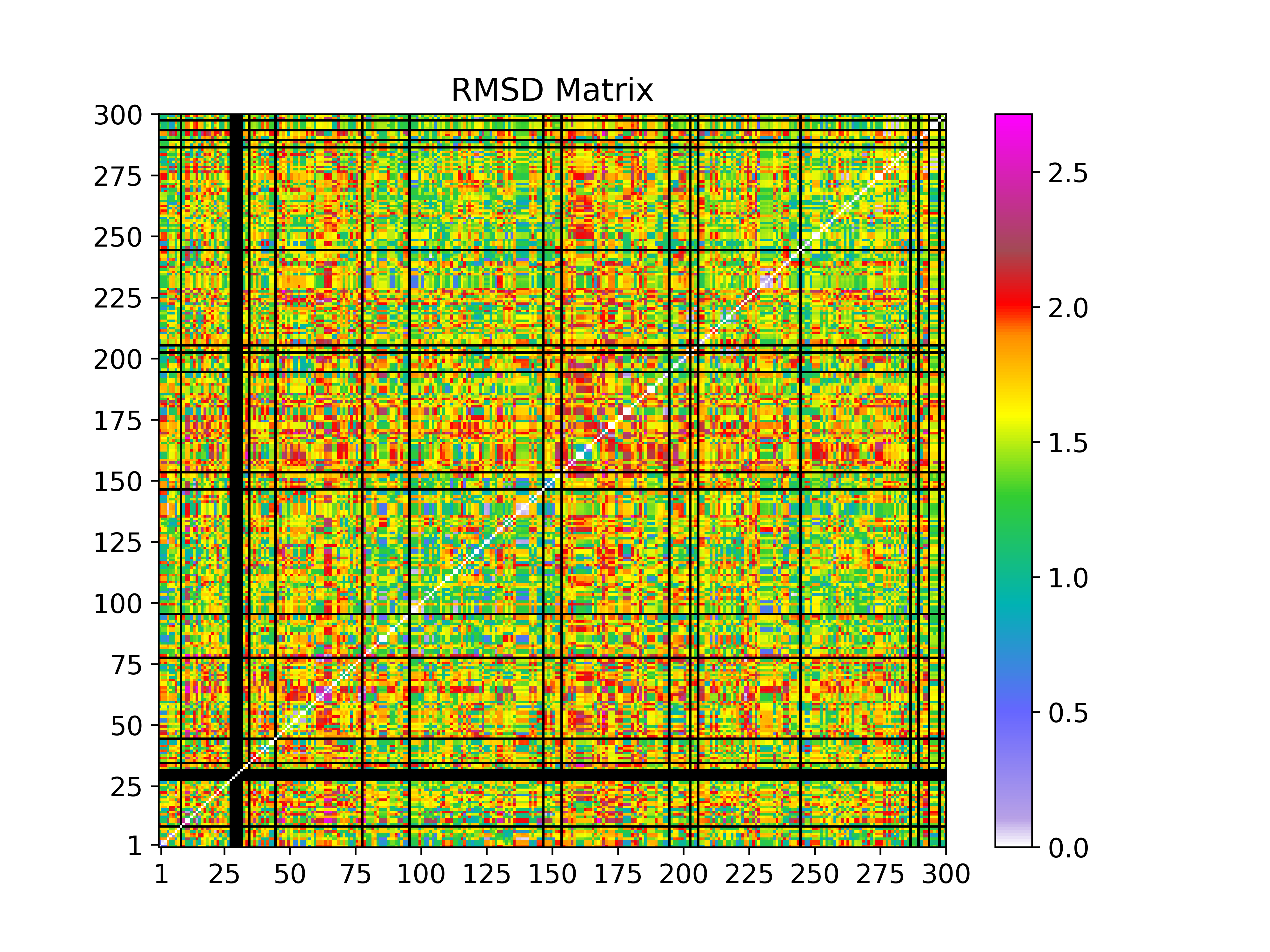}
    \caption{\label{NH3_RMSD_matrix}Full symmetric RMSD matrix results for the NH$_3$ binding behavior study on ASW ice model - comparison only performed between systems leading to accepted BE values.}
    \end{figure}

Fig. \ref{NH3_RMSD_matrix} presents the full RMSD analysis for NH$_3$ binding configuration sampling, in which each system associated to a valid BE (converged geometry optimization \& validation after check of imaginary frequency) is compared to each other valid system. Note that systems are labeled between 0 and 300, one cut being associated to 3 successive numbers accounting for its 3 geometries. 

\begin{figure}[]
    \centering
    \includegraphics[width=1\linewidth]{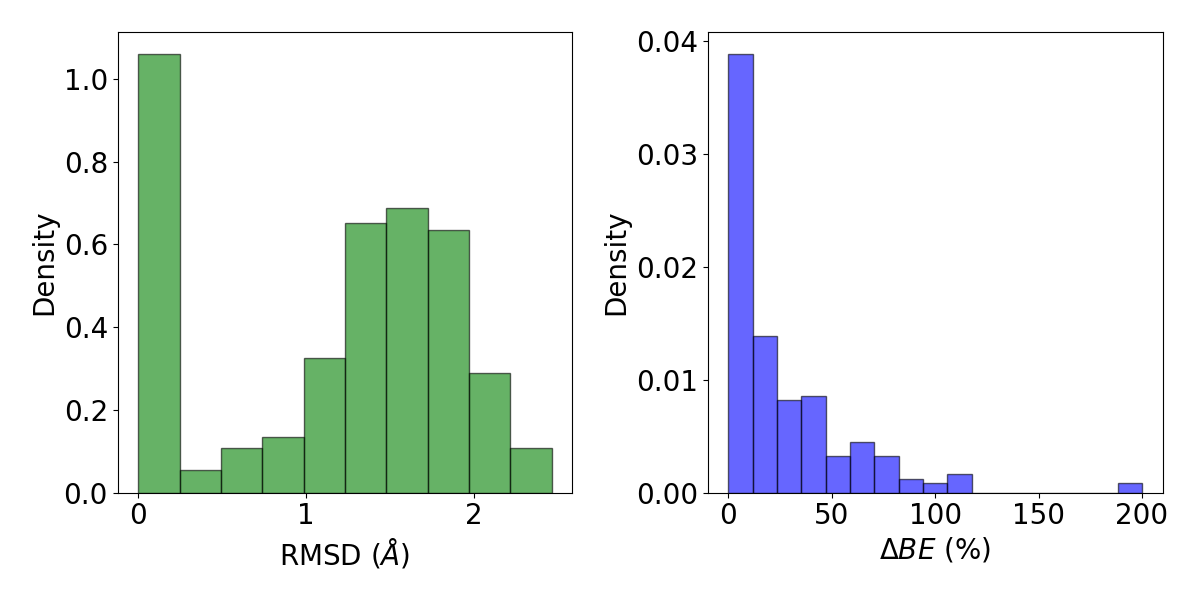}
    \caption{\label{NH3_histograms_RMSD_analysis_inter_geom_intra_cut}Histograms of (left) RMSD values, (right) $\Delta BE$ (\%) for comparison of optimized configurations from the different NH$_3$ starting orientations on the same cut.}
    \end{figure}

From these results, several statistical tests have been performed. Foremost, Fig. \ref{NH3_histograms_RMSD_analysis_inter_geom_intra_cut} gives the distributions of RMSD values and $\Delta BE$ for comparisons between optimized configurations of different NH$_3$ starting orientations on the same cut. From these plots, it is clear that the different starting orientations have an impact on the final complex geometry. This enforces the requirement of sampling these different orientations within a given cut to enable an extensive sampling of the diversity of binding configuration, including the local roughness of the substrate potential energy landscape. Pairs of geometries with RMSD value below 1 $\AA$ and $\Delta BE$ are disregarded in the in-site binding configurations redundancy analysis procedure, as discussed in the main text. These cutoffs are justified by the present analysis; indeed, integrating the contributions between 0 and 1 $\AA$ and 0 and \% respectively for the RMSD and $\Delta BE$ histograms of Fig. \ref{NH3_histograms_RMSD_analysis_inter_geom_intra_cut} yields consistent integrated densities, i.e. 0.29\% and 0.26\% respectively, supporting the coherence of the adopted thresholds.

\begin{figure}[]
    \centering
    \includegraphics[width=1\linewidth]{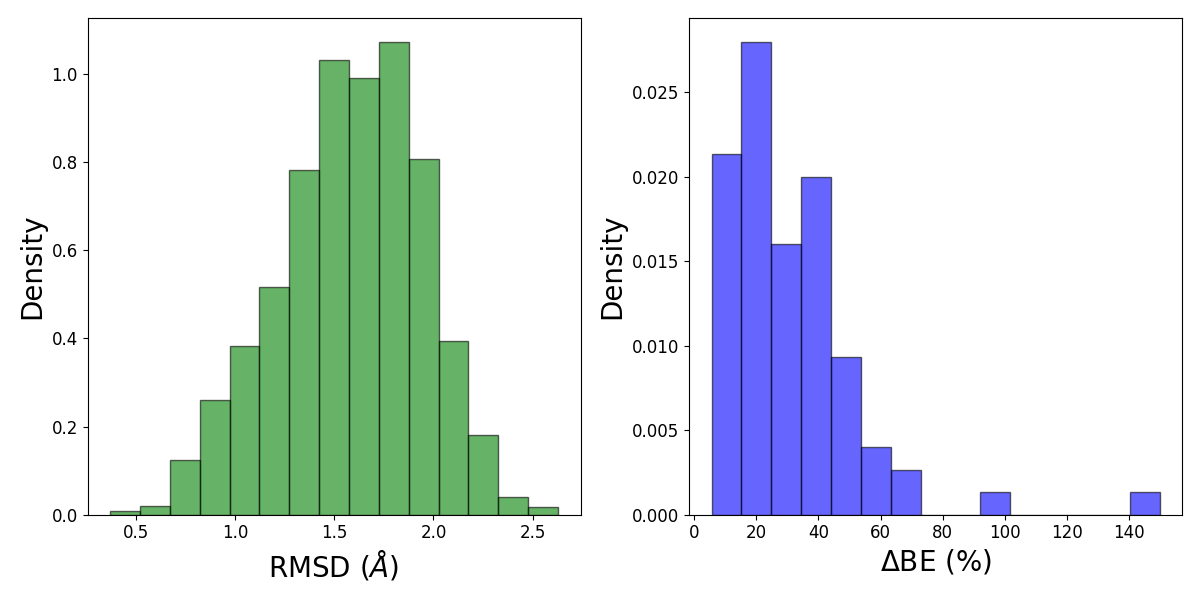}
    \caption{\label{NH3_histograms_RMSD_analysis_juxtaposed_systems}Histograms of (left) RMSD values, (right) $\Delta BE$ (\%) for comparisons between optimized configurations of juxtaposed cuts, including all NH$_3$ starting orientations.}
    \end{figure}

On the other hand, we also studied the RMSD value and corresponding $\Delta BE$ for the comparison of juxtaposed cuts, as shown in Fig. \ref{NH3_histograms_RMSD_analysis_juxtaposed_systems}. As we can see, the associated values are globally higher than 0.8 $\AA$, and $\Delta BE$ (\%) do not go below 5\% (min. 6.4). Concerning the full distribution of RMSD value including all the computed value except diagonal terms (comparison of systems with themselves, strictly leading to zero RMSD value) is characterized by a mean value of 1.56 $\pm$ 0.39 $\AA$. Excluding juxtaposed cuts as well as inter starting geometry intra cut contributions leads to a mean value of 1.56 $\pm$ 0.38 $\AA$. These results confirm that the grid spacing has been well chosen and that the first potential source  binding site sampling redundancy previously discussed has been effectively avoided. 

\subsection{CO related RMSD analyses}

\begin{figure}[]
    \centering
    \includegraphics[width=1\linewidth]{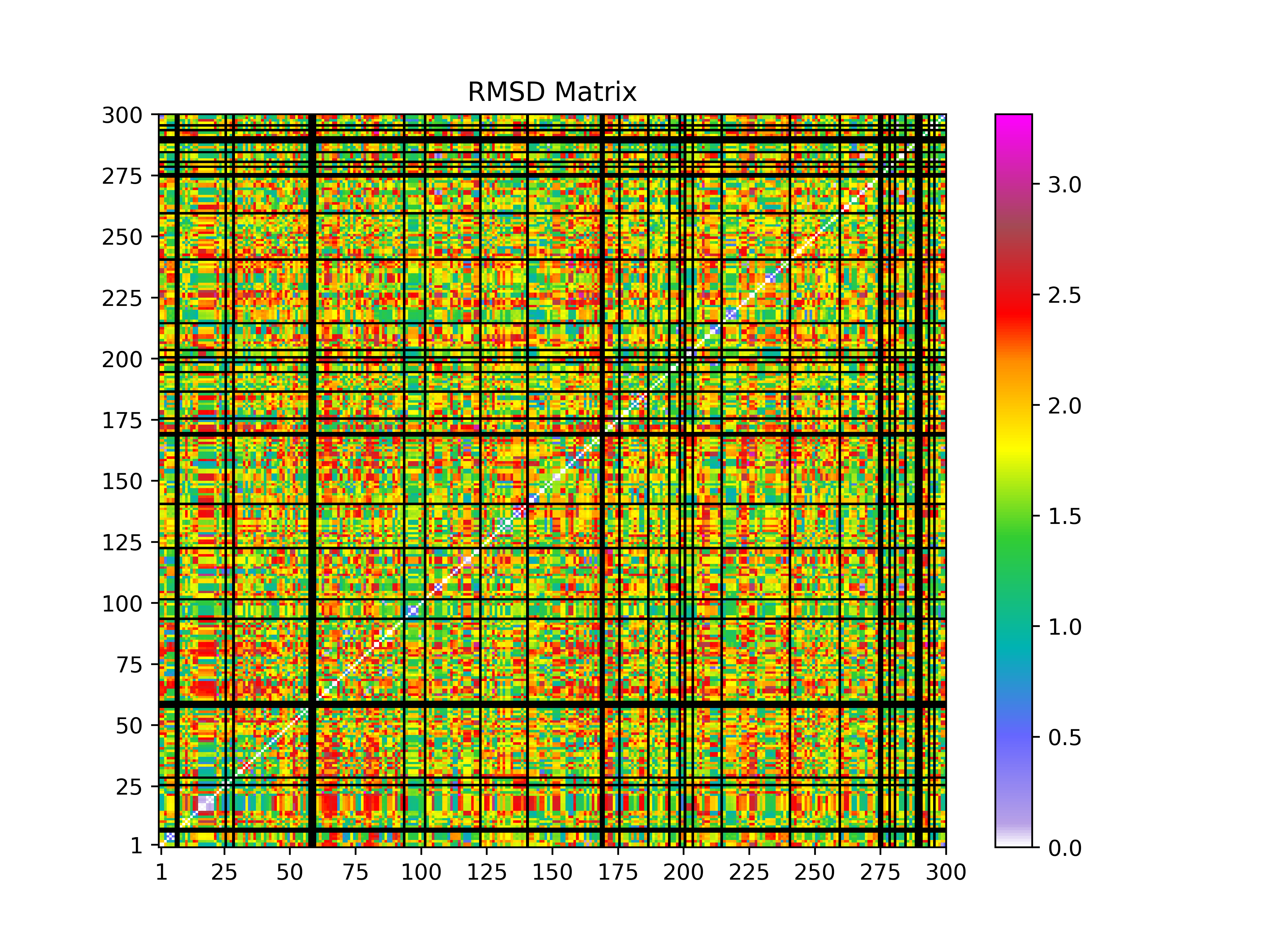}
    \caption{\label{CO_RMSD_matrix}Full symmetric RMSD matrix results for the CO binding behavior study on ASW ice model - comparison only performed between systems leading to accepted BE values.}
    \end{figure}

Fig. \ref{CO_RMSD_matrix} presents the full RMSD analysis for the binding configuration sampling of CO. 

    \begin{figure}[]
    \centering
    \includegraphics[width=1\linewidth]{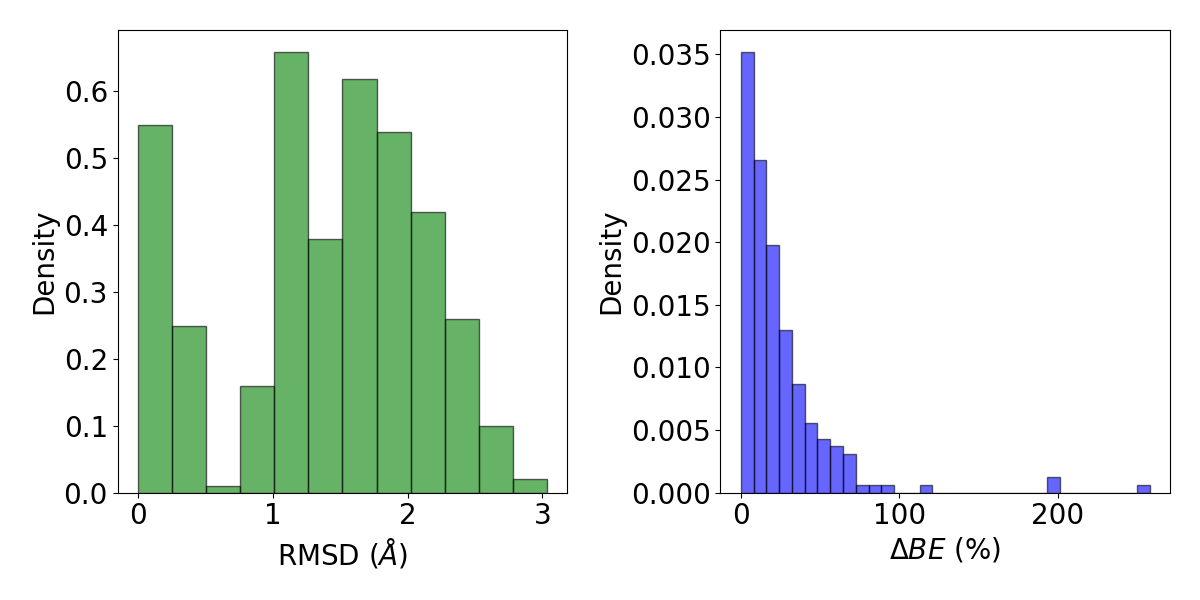}
    \caption{\label{CO_histograms_RMSD_analysis_inter_geom_intra_cut}Histograms of (left) RMSD values, (right) $\Delta BE$ for comparisons between optimized configurations of different CO starting orientations on the same cut.}
    \end{figure}

\begin{figure}[]
    \centering
    \includegraphics[width=1\linewidth]{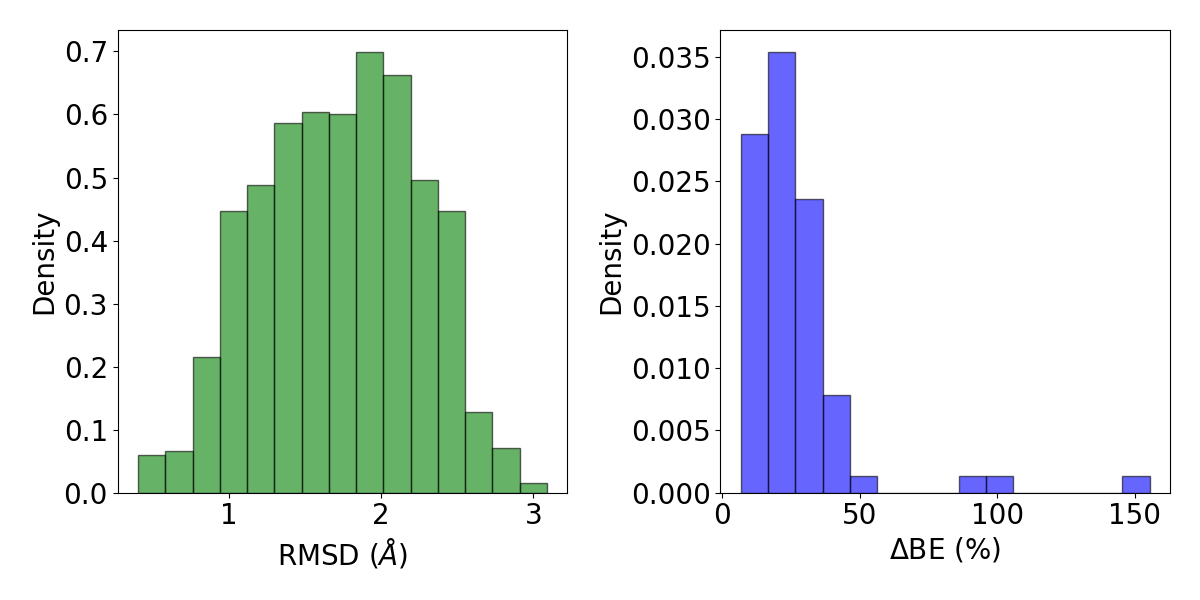}
    \caption{\label{CO_histograms_RMSD_analysis_juxtaposed_systems}Histograms of (left) RMSD values, (right) $\Delta BE$ for comparisons between optimized configurations of juxtaposed cuts, including all CO starting orientations.}
    \end{figure}

These same statistical tests as applied to NH$_3$ RMSD matrix have been applied, as shown in Figures \ref{CO_histograms_RMSD_analysis_inter_geom_intra_cut} and \ref{CO_histograms_RMSD_analysis_juxtaposed_systems} for the distributions of RMSD and $\Delta BE$ for the comparison respectively between optimized
configurations of different CO starting orientations on the same cut, and between juxtaposed cuts.

These two analyses lead to very similar qualitative trends to the case of NH$_3$. Once more, the coherence of the thresholds used for the definition of in-site redundancy is validated results in Fig.\ref{CO_histograms_RMSD_analysis_inter_geom_intra_cut}, demonstrating the consistency of the integrated densities between 0 and 1 $\AA$ and 0 and \% respectively for the RMSD (0.20\%) and $\Delta BE$ (0.18\%) histograms. Concerning the full RMSD distribution except diagonal terms, one gets a mean value of 1.78 $\pm$ 0.50 $\AA$. Excluding juxtaposed cuts as well as inter starting geometry intra cut contributions leads to a mean value of 1.79 $\pm$ 0.50 $\AA$. These results therefore also validate that the different starting orientations of the adsorbate onto a same system impact the resulting complex adsorbate-substrate geometry, while our spacing for our sampling grid is adequately chosen to avoid a collapse into the same energy minimum and therefore a redundancy between adjacent binding sites.

\subsection{CH$_4$ related RMSD analyses}

\begin{figure}[]
    \centering
    \includegraphics[width=1\linewidth]{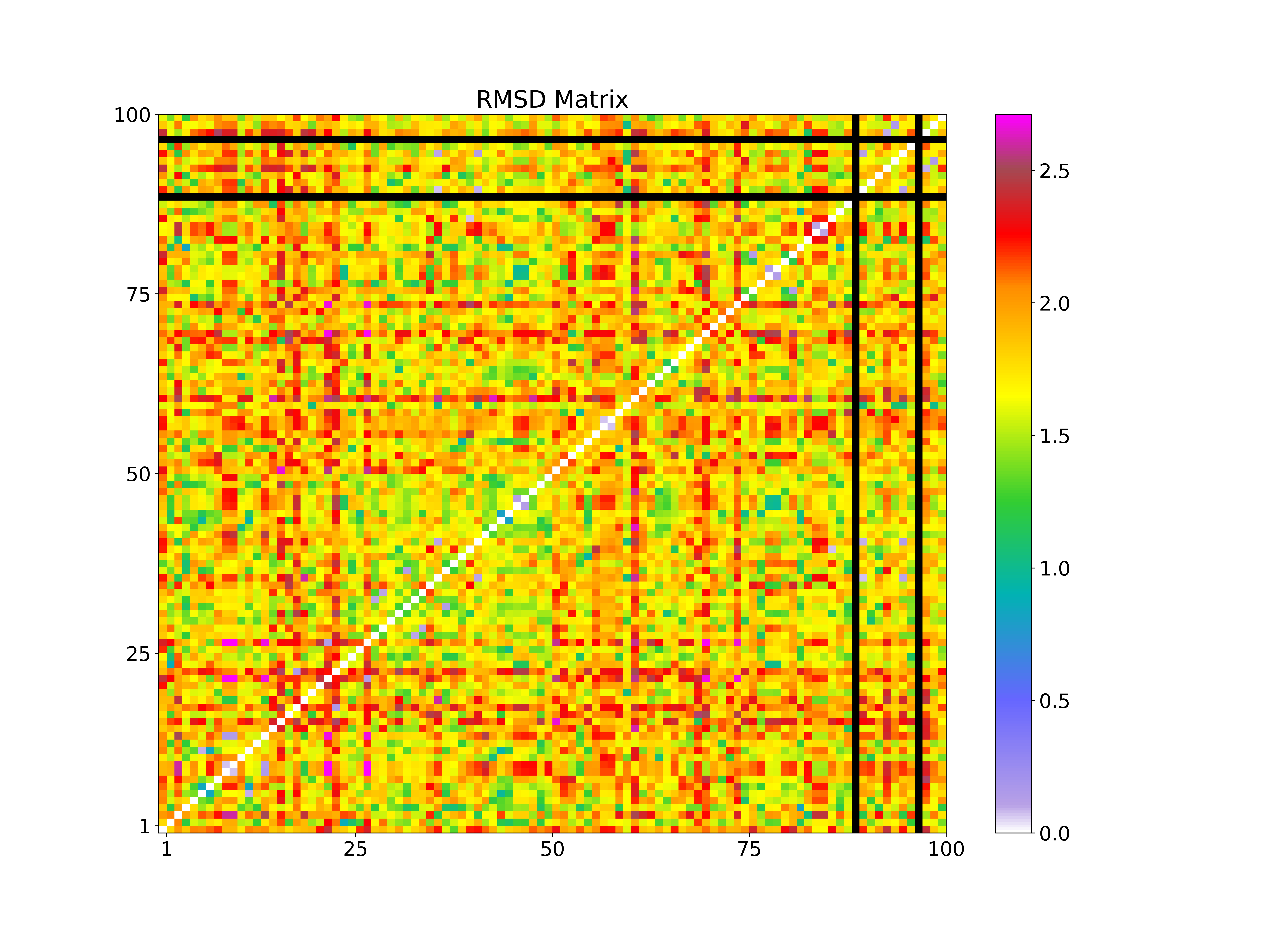}
    \caption{\label{CH4_RMSD_matrix}Full symmetric RMSD matrix results for the CH$_4$ binding behavior study on ASW ice model - comparison only performed between systems leading to accepted BE values.}
    \end{figure}

Remembering that the CH$_4$ binding behavior has been studied from a single starting geometry per hemisphere due to its symmetry, the RMSD matrix is composed of a 100 x 100 grid, given in Fig. \ref{CH4_RMSD_matrix}.

\begin{figure}[]
    \centering
    \includegraphics[width=1\linewidth]{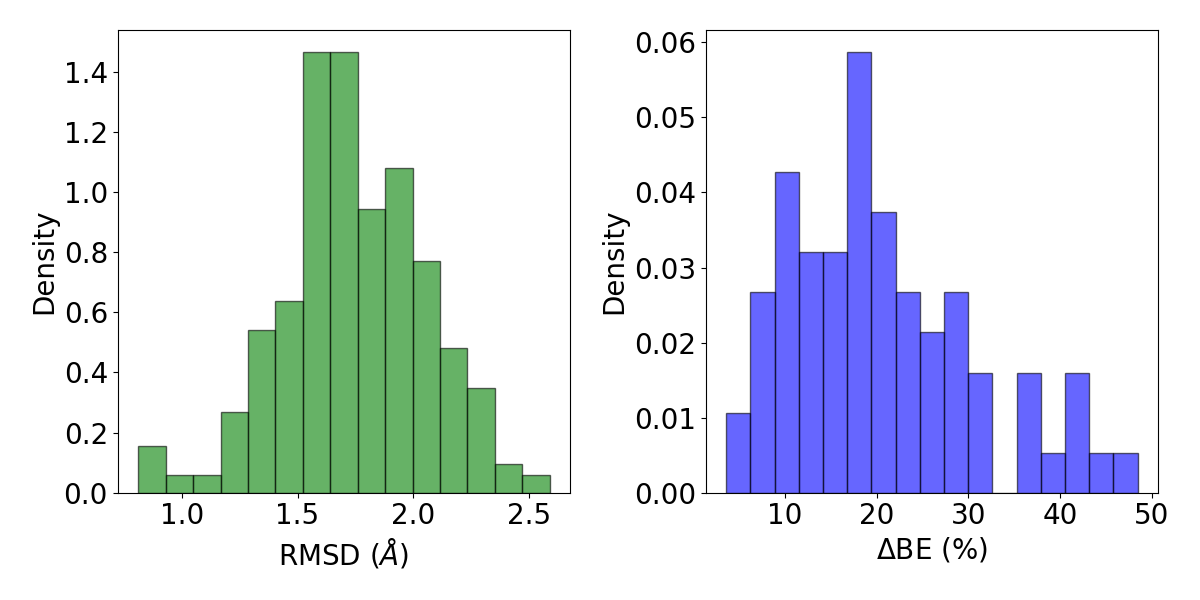}
    \caption{\label{CH4_histograms_RMSD_analysis_juxtaposed_systems}Histograms of (left) RMSD values, (right) $\Delta BE$ for comparisons between optimized configurations of juxtaposed cuts for CH$_4$ as adsorbate. }
    \end{figure}

The distributions of RMSD and $\Delta BE$ for the comparison between juxtaposed cuts is given in Figure \ref{CH4_histograms_RMSD_analysis_juxtaposed_systems}. Concerning the full
RMSD distribution except diagonal terms, one gets a mean value of 1.79 ± 0.31 $\AA$. Excluding juxtaposed cuts leads to a mean value of 1.78 ± 0.35 $\AA$. As for the two previously discussed adsorbate, juxtaposed cut does not lead to small RMSD associated to negligible $\Delta BE$ values, meaning that the initial grid spacing effectively avoids redundancy between juxtaposed binding sites. \bigskip

\end{appendix}

\end{document}